\documentclass{aa}
\usepackage[OT1]{fontenc}
\usepackage[latin1]{inputenc}
\usepackage{float}
\usepackage{amsmath}
\usepackage{graphicx}
\IfFileExists{url.sty}{\usepackage{url}}
                      {\newcommand{\url}{\texttt}}
\usepackage[authoryear]{natbib}

\makeatletter

\providecommand{\tabularnewline}{\\}

\usepackage{rotating}
\usepackage{afterpage}
\usepackage{hyperref}
\usepackage{txfonts}
\hypersetup{
pdfauthor = {Konrad R. W. Tristram},
pdftitle = {The nuclear dust torus in the Circinus galaxy}}
\usepackage[all]{hypcap}

\let\HyperRaiseLinkLength\@tempdima
\let\HyperRaiseLinkHook\@empty
\renewcommand{\HyperRaiseLinkDefault}{1.5\baselineskip}
\def\Hy@raisedlink#1{%
   \setlength\HyperRaiseLinkLength\HyperRaiseLinkDefault
   \HyperRaiseLinkHook
   \ifvmode
     #1%
   \else
     \penalty\@M
     \smash{\raise\HyperRaiseLinkLength\hbox{#1}}%
   \fi
 }

\authorrunning{Tristram et al.}
\titlerunning{The nuclear dust torus in the Circinus galaxy}

\abstract
{}
{To test the dust torus model for active galactic nuclei directly, we study the extent and morphology of the nuclear dust distribution in the \object{Circinus galaxy} using high resolution interferometric observations in the mid-infrared.}
{Observations were obtained with the MIDI instrument at the Very Large Telescope Interferometer. The 21 visibility points recorded are dispersed with a spectral resolution of $\lambda/\delta\lambda\approx30$ in the wavelength range from $8$ to $13\,\mu\mathrm{m}$. To interpret the data we used a stepwise approach of modelling with increasing complexity. The final model consists of two black body Gaussian distributions with dust extinction.}
{We find that the dust distribution in the nucleus of Circinus can be explained by two components, a dense and warm disk-like component of  $0.4\,\mathrm{pc}$ size and a slightly cooler, geometrically thick torus component with a size of $2.0\,\mathrm{pc}$. The disk component is oriented perpendicular to the ionisation cone and outflow and seems to show the silicate feature at $10\,\mu\mathrm{m}$ in emission. It coincides with a nuclear maser disk in orientation and size. From the energy needed to heat the dust, we infer a luminosity of the accretion disk of $L_{\mathrm{acc}}=10^{10}\,\mathrm{L}_{\odot}$, which corresponds to $20\,\%$ of the Eddington luminosity of the nuclear black hole. We find that the interferometric data are inconsistent with a simple, smooth and axisymmetric dust emission. The irregular behaviour of the visibilities and the shallow decrease of the dust temperature with radius provide strong evidence for a clumpy or filamentary dust structure. We see no evidence for dust reprocessing, as the silicate absorption profile is consistent with that of standard galactic dust. We argue that the collimation of the ionising radiation must originate in the geometrically thick torus component.}
{Based on a great leap forward in the quality and quantity of interferometric data, our findings confirm the presence of a geometrically thick, torus-like dust distribution in the nucleus of Circinus, as required in unified schemes of Seyfert galaxies. Several aspects of our data require that this torus is irregular, or ``clumpy''.}


\makeatother
\begin{document}

\title{Resolving the complex structure of the dust torus in the active nucleus
of the Circinus galaxy\thanks{Based on observations collected at the European Southern Observatory, Chile, program numbers 073.A-9002(A), 060.A-9224(A), 074.B-0213(A/B), 075.B-0215(A) and 077.B-0026(A).}}

\author{K. R. W. Tristram\inst{1}, K. Meisenheimer\inst{1}, W. Jaffe\inst{2},
M. Schartmann\inst{1}, H.-W. Rix\inst{1}, Ch. Leinert\inst{1}, S. Morel\inst{3}, \\
M. Wittkowski\inst{4}, H. Röttgering\inst{2}, G. Perrin\inst{5}, B. Lopez\inst{6},
D. Raban\inst{2}, W. D. Cotton\inst{7}, U. Graser\inst{1}, F. Paresce\inst{4}, \\
Th. Henning\inst{1}}

\institute{Max-Planck-Institut für Astronomie, Königstuhl 17, 69117 Heidelberg,
Germany \and Leiden Observatory, Leiden University, Niels-Bohr-Weg
2, 2300 CA Leiden, The Netherlands \and European Southern Observatory,
Casilla 19001, Santiago 19, Chile \and European Southern Observatory,
Karl-Schwarzschild-Strasse 2, 85748 Garching bei München, Germany
\and LESIA, UMR 8109, Observatoire de Paris-Meudon, 5 place Jules
Janssen, 92195 Meudon Cedex, France  \and Laboratoire Gemini, UMR
6203, Observatoire de la Côte d'Azur, BP 4229, 06304 Nice Cedex 4,
France \and NRAO, 520 Edgemont Road, Charlottesville, VA 22903-2475,
USA}

\mail{tristram@mpia.de}

\date{Received 27 July 2007 / 31 August 2007}

\maketitle

\section{Introduction\label{sec:circ_introduction}}

In the widely accepted model for low and intermediate luminosity active
galactic nuclei (AGN), the central engine, consisting of a hot accretion
disk around a supermassive black hole, and the broad line region (BLR)
are surrounded by a torus of obscuring dust. Where the radiation from
the central engine can escape, an ionisation cone and the narrow line
region (NLR) are formed. Although motivated by the observation of
broad emission lines in polarised light in type 2 AGN (torus edge-on),
most tests of this {}``unified model'' rely on the shape of the
overall spectral energy distribution (SED) and statistics, \emph{i.e.}
the number of objects where the central source is obscured or visible.
Until recently, a direct spatially resolved assessment of the dust
distribution that allegedly is responsible for the anisotropic absorption
and obscuration has been missing. It is too small to be significantly
resolved with single dish telescopes in the infrared, even for $10\,\mathrm{m}$
class telescopes at the diffraction limit. It has hence been impossible
to directly confirm or disprove the existence and geometry of this
essential component. Properties of the dust distribution, such as
its size, temperature, geometry, alignment, and filling factor have
hence remained unknown. A new approach is possible by studying the
nearest AGN using interferometry at infrared wavelengths. The first
objects to be successfully analysed in this way are the Seyfert 1
galaxy \object{NGC~4151} \citep{2003Swain} and the Seyfert 2 galaxy
NGC~1068 \citep{2004Wittkowski,2004Jaffe1,2006Poncelet}. In \object{NGC~1068},
which is the brightest galaxy in the mid-infrared (MIR), indeed an
extended dust structure was resolved. In this paper we study the second
brightest AGN in the MIR, the Circinus galaxy.

The Circinus galaxy is a highly inclined ($\sim65^\circ$) SA(s)b galaxy
harbouring a Seyfert type 2 active nucleus as well as a nuclear starburst.
At about 4 Mpc distance ($1\,\mathrm{arcsec}\sim20\,\mathrm{pc}$,
\citealt{1977Freeman}), the galaxy is among the nearest AGN and hence
it is an ideal object to study the nuclear region of active galaxies.
The nucleus is heavily obscured by dust lanes in the plane of the
galaxy, so that it is best observed in the infrared. The galaxy can
be considered a prototype Seyfert 2 object due to the presence of
narrow emission lines \citep{1994Oliva}, broad emission lines in
polarised light \citep{1998Oliva}, an ionisation cone traced by {[}\ion{O}{iii}{]},
H$\alpha$ and {[}\ion{Si}{vii}{]} \citep{1997Veilleux,2000Maiolino,2000Wilson,2004Prieto},
an outflow observed in CO \citep{1998Curran,1999Curran}, bipolar
radio lobes \citep{1998Elmouttie} and a Compton reflection component
in X-rays \citep{1996Matt,2001Smith}. The ionisation cone, the outflow
and the radio lobes have a position angle of about $-40^\circ$, which
is roughly perpendicular to the disk of the galaxy, to the circum-nuclear
star forming rings and to the nuclear rings of molecular gas. \citet{2003Greenhill}
have found an edge-on, warped {}``accretion'' disk with a radius
of $r_{\mathrm{outer}}\sim0.40\,\mathrm{pc}$ which is traced by $\mathrm{H}_{2}\mathrm{O}$
masers. The almost Keplerian velocity curve yields an estimate of
the mass of the central black hole to $M_{\mathrm{BH}}<(1.7\pm0.3)\cdot10^{6}\,\mathrm{M}_{\odot}$.

Modelling the SED of the nucleus of the Circinus galaxy was attempted
by \citet{1997Siebenmorgen}, \citet{2001Ruiz} and \citet{2005Schartmann}.
The latter two came to the conclusion that the presumed toroidal dust
distribution is relatively small, with an effective emission region
in the MIR of less than $3\,\mathrm{pc}$ (\emph{i.e.} $<0.15\,\mathrm{arcsec}$).

A detailed observational analysis in the mid-infrared was performed
by \citet{2005Packham}. The authors found extended emission with
$2\,\mathrm{arcsec}$ size at position angles of $81^\circ$ and $278^\circ$,
which is consistent with the edges of the aforementioned ionisation
cones. However, their point spread function (PSF) analysis indicated
that any extended nuclear emission in the mid-infrared has sizes less
than $0.20\,\mathrm{arcsec}$ ($4\,\mathrm{pc}$) or is very weak.
This showed again that single dish observations with current telescopes
are not capable of resolving the nuclear dust distribution even for
the nearest AGN, such as Circinus. For this reason, interferometric
observations of the Circinus nucleus in the mid-infrared with the
Very Large Telescope Interferometer (VLTI) were performed.

This paper is organised as follows. In Sect.~\ref{sec:circ_observ-and-reduc}
we present the details of the observations and of the data reduction.
Sect.~\ref{sec:circ_results} describes the properties of the interferometric
data and the results directly deducible from them. Sect.~\ref{sec:circ_modelling}
treats the modelling of the data. The interpretation of our findings
and implications thereof are discussed in Sect.~\ref{sec:circ_discussion}.
Our conclusions are given in Sect.~\ref{sec:circ_conclusions}.

\section{Observations and data reduction\label{sec:circ_observ-and-reduc}}

\subsection{Instrument\label{sub:circ_instrument}}

We obtained interferometric observations with the MID-infrared Interferometric
instrument (MIDI) at the VLTI located on Cerro Paranal in northern
Chile and operated by the European Southern Observatory (ESO). MIDI
is a two beam Michelson type interferometer producing dispersed fringes
in the N band in a wavelength range from $8$ to $13\,\mu\mathrm{m}$
\citep{2003Leinert2}. For our observations of Circinus, we combined
the light of two of the $8.2\,\mathrm{m}$ unit telescopes (UTs) at
any one time and used the low spectral resolution mode ($\lambda/\delta\lambda\approx30$)
by insertion of a NaCl prism into the light path after the beam combiner.

\subsection{Observations\label{sub:circ_observations}}

The data were acquired between February 2004 and May 2006 during a
total of six epochs. The first data set was obtained in February 2004
in science demonstration time (SDT). All following observations were
carried out using guaranteed time observations (GTO). Table~\ref{table:circ_observation-log}
gives a summary of the observations.

\begin{sidewaystable*}

\caption{Log of the observations, including the detector integration times
(DIT), the number of frames (NDIT) for the fringe tracks and for the
photometries as well as the lengths (BL) and the position angles (PA)
of the projected baselines. Times, ambient values and baseline properties
are for the start of the fringe track.}

\label{table:circ_observation-log}\renewcommand{\footnoterule}{}

\begin{tabular}{ccccccccccll}
\hline 
\hline Date and time&
\multicolumn{2}{c}{Object}&
DIT&
\multicolumn{2}{c}{NDIT}&
Airmass&
Seeing%
\footnote{From the seeing monitor (DIMM).%
}&
BL&
PA&
\multicolumn{2}{l}{Associated calibrator and comments}\tabularnewline
{[}UTC{]}&
&
&
{[}s{]}&
fringes&
phot.%
\footnote{A missing value signifies that no photometry was observed. A value
in brackets denotes that the photometry was unusable.%
}&
&
{[}$\mathrm{arcsec}${]}&
{[}m{]}&
{[}°{]}&
&
\tabularnewline
\hline
\textbf{2004 Feb 12: }&
\textbf{UT3~--~UT2}&
&
&
&
&
&
&
&
&
&
\tabularnewline
\hline
06:55:11&
Circinus&
sci01&
0.020&
12000&
(2000)&
1.473&
0.96&
43.49&
19.30&
cal01&
photometry unusable, used phot of sci02 \tabularnewline
07:06:46&
Circinus&
sci02&
0.020&
12000&
2000&
1.449&
0.91&
43.36&
21.49&
cal02&
photometry used for sci01 and sci02\tabularnewline
07:36:34&
HD\,120404&
cal01&
0.012&
12000&
2000&
1.456&
1.08&
41.39&
32.16&
&
\tabularnewline
08:27:38&
HD\,120404&
cal02&
0.012&
12000&
2000&
1.416&
0.56&
40.33&
42.18&
&
\tabularnewline
\textbf{2004 Jun 03:}&
\textbf{UT3~\--~UT2}&
&
&
&
&
&
&
&
&
&
\tabularnewline
\hline
05:50:45&
Circinus&
sci03&
0.012&
8000&
1500&
1.693&
0.68&
29.16&
92.99&
cal03&
\tabularnewline
06:42:24&
HD\,120404&
cal03&
0.012&
8000&
1500&
2.095&
0.65&
25.40&
117.87&
&
\tabularnewline
07:58:04&
Circinus&
sci04&
0.012&
8000&
(1500)&
2.542&
0.73&
20.74&
129.68&
cal03&
phot B unusable, replaced by copy of phot A\tabularnewline
\textbf{2005 Feb 21:}&
\textbf{UT2~--~UT4}&
&
&
&
&
&
&
&
&
&
\tabularnewline
\hline
04:38:01&
HD\,107446&
cal05&
0.012&
8000&
1500&
1.367&
0.83&
87.87&
51.88&
&
\tabularnewline
05:25:00&
Circinus&
sci05&
0.012&
8000&
&
1.617&
0.77&
87.38&
35.68&
cal05&
use photometry from 2005-03-01T04:38:30\tabularnewline
\textbf{2005 Mar 01:}&
\textbf{UT3~--~UT4}&
&
&
&
&
&
&
&
&
&
\tabularnewline
\hline
03:34:58&
HD\,120404&
cal06&
0.012&
8000&
1500&
1.906&
0.95&
50.87&
39.76&
&
$\mathtt{dsky=-2}$ (see Sect.~\ref{sub:circ_data-reduction})\tabularnewline
04:06:04&
Circinus&
sci06&
0.012&
8000&
1500&
1.809&
0.75&
49.33&
44.58&
cal06&
$\mathtt{dsky=-2}$ (see Sect.~\ref{sub:circ_data-reduction})\tabularnewline
04:38:30&
Circinus&
sci07&
0.012&
8000&
1500&
1.671&
0.72&
50.99&
53.61&
cal07&
photometry used for sci07 and sci08\tabularnewline
04:49:21&
Circinus&
sci08&
0.012&
8000&
&
1.631&
0.71&
51.57&
56.54&
cal07&
no photometry observed, used phot of sci07\tabularnewline
05:12:31&
HD\,120404&
cal07&
0.012&
8000&
1500&
1.581&
0.55&
54.82&
65.93&
&
\tabularnewline
06:58:52&
Circinus&
sci09&
0.012&
8000&
1500&
1.358&
0.74&
58.09&
88.34&
cal08&
\tabularnewline
07:28:53&
HD\,120404&
cal08&
0.012&
8000&
1500&
1.412&
0.48&
59.77&
98.71&
&
\tabularnewline
09:21:08&
Circinus&
sci10&
0.012&
8000&
1500&
1.345&
0.64&
61.95&
120.00&
cal09&
photometry used for sci10 and sci11\tabularnewline
09:43:25&
HD\,120404&
cal09&
0.012&
8000&
1500&
1.480&
0.78&
62.10&
129.42&
&
\tabularnewline
10:08:22&
Circinus&
sci11&
0.012&
8000&
&
1.400&
1.25&
62.36&
130.50&
cal09&
no photometry observed, used phot of sci10\tabularnewline
\textbf{2005 May 26:}&
\textbf{UT2~--~UT3}&
&
&
&
&
&
&
&
&
&
\tabularnewline
\hline
22:57:46&
HD\,120404&
cal10&
0.012&
8000&
1500&
1.666&
0.60&
42.62&
11.21&
&
\tabularnewline
23:29:03&
Circinus&
sci12&
0.012&
8000&
3000&
1.563&
0.63&
43.77&
12.79&
cal10&
reset chopping in second half of phot B\tabularnewline
23:43:07&
Circinus&
sci13&
0.012&
5000&
3000&
1.524&
0.54&
43.67&
15.47&
cal10&
\tabularnewline
01:37:29&
Circinus&
sci14&
0.018&
5000&
3000&
1.340&
0.67&
41.98&
37.05&
cal11&
\tabularnewline
02:02:11&
HD\,120404&
cal11&
0.018&
5000&
1500&
1.409&
1.52&
39.64&
47.39&
&
\tabularnewline
03:40:07&
HD\,120404&
cal12&
0.018&
5000&
1500&
1.454&
0.85&
36.24&
67.11&
&
\tabularnewline
04:07:02&
Circinus&
sci15&
0.018&
5000&
3000&
1.374&
0.75&
36.82&
65.28&
cal12&
\tabularnewline
04:17:27&
Circinus&
sci16&
0.018&
5000&
3000&
1.387&
0.98&
36.31&
67.31&
cal12&
\tabularnewline
04:28:22&
Circinus&
sci17&
0.018&
5000&
3000&
1.404&
1.05&
35.77&
69.45&
cal12&
\tabularnewline
\textbf{2006 May 18:}&
\textbf{UT2~--~UT3}&
&
&
&
&
&
&
&
&
&
\tabularnewline
\hline
06:16:21&
Circinus&
sci18&
0.018&
8000&
4000&
1.560&
1.50&
31.69&
84.18&
cal13&
\tabularnewline
06:42:24&
HD\,120404&
cal13&
0.018&
8000&
4000&
1.788&
0.84&
28.96&
100.07&
&
\tabularnewline
07:09:48&
Circinus&
sci19&
0.018&
8000&
4000&
1.754&
0.98&
28.21&
96.40&
cal13&
\tabularnewline
07:36:18&
HD\,120404&
cal14&
0.018&
8000&
4000&
2.035&
0.78&
25.98&
114.66&
&
rejected due to very low correlated flux\tabularnewline
08:04:13&
Circinus&
sci20&
0.018&
8000&
4000&
2.049&
0.59&
24.51&
110.75&
cal13&
photometry used for sci20 and sci21\tabularnewline
08:17:34&
Circinus&
sci21&
0.018&
8000&
&
2.142&
0.67&
23.61&
114.69&
cal13&
no photometry observed, used phot of sci20\tabularnewline
\hline
\end{tabular}\normalsize

\end{sidewaystable*}

The observing sequence repeated for the science target and calibrator
star consists of the following steps:

First, acquisition images of the two beams A and B coming from the
unit telescopes are taken simultaneously. For this purpose, the standard
chopping technique is used to suppress the background in the mid-infrared.
For our observations, we used the default values for the chopping
frequency ($f=2\,\mathrm{Hz}$), chopping angle ($\alpha=0$) and
chopping throw ($\delta=15\,\mathrm{arcsec}$) except for the observations
on 2004 Jun 03, where a smaller throw of $\delta=10\,\mathrm{arcsec}$
was used. For the acquisition of Circinus, we used the SiC filter
($11.8\pm2.3\,\mu\mathrm{m}$) at longer wavelengths because of the
rising spectrum of the source. For the acquisition of the calibrator
stars, which are in the Rayleigh-Jeans part of their spectra, we used
the N8.7 filter ($8.6\pm1.5\,\mu\mathrm{m}$) at shorter wavelengths.
The acquisition procedure was repeated until the overlap of the two
beams was within one pixel accuracy, that is within $\sim86\,\mathrm{mas}$.

Then the beam combiner, the $0.6\,\mathrm{arcsec}$ wide slit and
the NaCl prism are inserted into the light path leading to two spectrally
dispersed interferograms of opposite phase. No chopping is needed
during the interferometric measurement as the background is uncorrelated
and cancels out in the data reduction process. The fringe signal is
searched by scanning the optical path difference (OPD) over a few
mm at the expected position. This is achieved by adjusting the VLTI
delay lines. At the same time, a piezo-driven mirror in MIDI performs
a sawtooth scanning pattern with a scan length of $80\,\mu\mathrm{m}$
for one of the delay lines. This modulation generates moving fringes
from which the optical path difference can be estimated. Once the
zero optical path difference (0~OPD) is found, the so-called fringe
track is started, during which the path difference is stabilised using
the scanning of the piezo driven mirrors, while the VLTI delay lines
compensate for siderial motion and atmospheric OPD shifts. There are
two possibilities to track the fringes: the scans can pass over the
white light fringe at 0~OPD or they are performed next to it. We
preferred to observe in {}``offset tracking'' mode with the OPD
stabilised at $50\,\mu\mathrm{m}$ from 0~OPD, because this is of
advantage for a coherent data analysis (see below). The observations
on 2004 Jun 03 and 2005 Feb 21 and the first visibility point on 2005
Mar 01 were, however, observed tracking on the white light fringe.
During these periods, the standard tracking software implemented at
Paranal did not allow for using offset tracking.

Finally, photometric data are recorded using only one telescope at
a time in the otherwise identical optical set-up. This results in
two photometric data sets, one for each telescope. Again, chopping
has to be used to suppress the background in the mid-infrared.

Because our data were obtained over such a long period of time, the
observation techniques evolved in the meantime. Additionally, the
instrument itself has improved considerably since the first observations
were made in February 2004. This leads to a large variety of exposure
times ranging from $\mathit{DIT}=12\,\mathrm{ms}$ to $20\,\mathrm{ms}$.
Similarly, the frame numbers range from $\mathit{NDIT}=5000$ to $12000$
for the fringe tracking and from $\mathit{NDIT}=1500$ to $4000$
for the photometry (see Table \ref{table:circ_observation-log}).
For the first 4 visibility points, only a tip tilt correction for
the incoming wavefronts was possible using the STRAP units \citep{2000Bonaccini}.
A much better correction was obtained with the MACAO adaptive optics
system \citep{2003Arsenault} since its implementation in late 2004,
\emph{i.e.} starting with our observations in February 2005. For Circinus,
both systems were operated with a separate guide star at $50\pm2\,\mathrm{arcsec}$
distance from the nucleus. For the observations of the calibrators,
the wavefront sensing was performed using the stars themselves. The
overall ambient conditions for all the observations were good to fair
with seeing values from the site monitor varying from $0.5$ to $1.8\,\mathrm{arcsec}$.
We had no problems with clouds or constraints due to wind, except
for one observation. This observation ({}``sci05''), carried out
in service mode on the UT2~--~UT4 baseline, had to be aborted due
to clouds building up over Paranal. Thus, no photometric data could
be obtained for this measurement.

\subsection{Data reduction\label{sub:circ_data-reduction}}

Not all measurements contain the full data set consisting of a fringe
track and two photometric measurements, so that they could be reduced
in a straightforward manner. This has three reasons: first of all,
several measurements were performed at the end of the night or of
the allocated time and only the fringe track could be obtained. Secondly,
for some measurements the photometries proved to be unusable or to
be of very low quality. Finally, for the measurement on the UT2~--~UT4
baseline, no photometry could be obtained at all due to the adverse
weather conditions. To nevertheless make use of all the fringe tracks
obtained, we completed the fragmentary data sets with photometric
data from the other measurements preferably using measurements as
close as possible in time. The details are included in the last column
of Table \ref{table:circ_observation-log}.

The data reduction was entirely performed by coherent visibility estimation
using the software package EWS (Expert Work Station), written by Walter
Jaffe. The alternative software package MIDI Interactive Analysis
(MIA), which is based on a power spectrum analysis, was only used
to cross-check the results for a selected number of observations:
in these cases the two methods gave consistent results%
\footnote{An online manual of both MIA and EWS is available at \url{http://www.strw.leidenuniv.nl/~koehler/MIA+EWS-Manual/}.%
}. A detailed description of the coherent method can be found in \citet{2004Jaffe2}.
Here, we will concentrate only on the essential steps and highlight
our specific configuration during the data reduction:

First, a fixed mask is applied to the two opposite phased interferograms,
which are then collapsed to form two one-dimensional interferograms.
These are then subtracted with the result of doubling the signal while
removing most of the background. As the fringes vary rapidly in time
due to the scanning of the MIDI internal mirrors, a high-pass filter
in temporal direction with a width of, in our case, 10 frames ($\mathtt{smooth=10}$)
is applied to further remove the background. Additionally, the average
of the entire spectrum is subtracted ($\mathtt{dave=1}$), assuming
a modulation of the fringe signal in frequency space. This is indeed
the case, when the OPD is not zero, hence our preference to track
next to the white light fringe. In the following two steps, the known
instrumental OPD and the atmospheric delay are removed. The latter
is estimated from the group delay ($\mathtt{gsmooth=10}$). Additionally
a constant phase shift induced by the varying index of refraction
of water vapour is estimated and also removed. After rejecting frames
with too low or too high optical path difference, the remaining data
is averaged coherently yielding the raw correlated flux, $\mathsf{rawcor}$,
and the differential phase, that is the relative variations of the
phase with wavelength.

To determine the raw total flux of the target, first standard data
reduction methods for chopped data are applied to the photometric
data: the sky frames are subtracted from the object frames. As this
does not lead to a satisfactory sky subtraction, two stripes on the
detector frame, one above the target position and one below, are used
to subtract an additional wavelength dependent sky estimate. Only
for one data set an adjustment of the position of these stripes using
the option $\mathtt{dsky=-2}$ was necessary (see the respective comment
in Table \ref{table:circ_observation-log}). The data are then compressed
and averaged to a one-dimensional photometry, one for every telescope
and side of the beam splitter ({}``channel''). Two sets of total
fluxes are extracted: One is needed for the determination of the visibility,
and hence the same mask as for the interferometric data is applied.
To account for the possible inequality of the two beams from the two
telescopes, this total flux is calculated as the geometric mean: $\mathsf{rawtotg}=\sqrt{A_{1}\cdot B_{1}}+\sqrt{A_{2}\cdot B_{2}}$,
with $A_{1}$ the photometry of beam A (from the first telescope)
in channel 1, $B_{1}$ the photometry of beam B (from the second telescope)
in channel 1, as well as $A_{2}$ and $B_{2}$ the corresponding beams
in channel 2. The second total flux is determined without applying
a mask and by taking the arithmetic mean of the individual measurements:
$\mathsf{rawtota}=(A_{1}+A_{2}+B_{1}+B_{2})/4$. The raw visibility
is defined as $\mathsf{rawvis}=\mathsf{rawcor}/\mathsf{rawtotg}$.
The same data reduction is applied to both the science target and
to the calibrator star leading to the raw correlated flux $\mathsf{scirawcor}$
and $\mathsf{calrawcor}$ respectively, as well as the raw total fluxes
$\mathsf{scirawtotg}$, $\mathsf{calrawtotg}$, $\mathsf{scirawtota}$,
$\mathsf{calrawtota}$ and the raw visibilities $\mathsf{scirawvis}$
and $\mathsf{calrawvis}$.

Finally, the raw fluxes and visibilities of the science target are
calibrated using the data of the calibrator star, to account for instrumental
visibility losses and to determine the absolute flux values. For the
calibration, we exclusively used the calibrator HD\,120404 associated
with Circinus, with the exception of one case, the long baseline UT2~--~UT4,
where HD\,120404 was not observed at all. In this case, we used HD\,107446
instead. The quality of the calibrators were checked by comparison
to all the other calibrators observed during the same night. We decided
to reject only one calibrator measurement because of its low correlated
flux compared to the other measurements during the night: {}``cal14''
observed on 2006 May 18 at 07:36:18. We corrected the Rayleigh-Jeans
approximation for the template spectrum of the calibrator using a
template spectrum, $\mathsf{caltemplate}$, from a database provided
by Roy van Boekel, (\emph{private comm}.). These template spectra
were obtained by fitting stellar templates to multiband photometry.

To keep the influence of errors in the photometry as small as possible
(especially considering the smaller number of usable photometries
with respect to the number of successful fringe tracks), we calculated
the calibrated correlated flux of Circinus according to $F_{\mathrm{cor}}(\lambda)=\mathsf{scicalcor}=\mathsf{scirawcor}/\mathsf{calrawcor}\cdot\mathsf{caltemplate}$
using the option $\mathtt{nophot=1}$ in EWS. The calibrated total
flux is given by $F_{\mathrm{tot}}(\lambda)=\mathsf{scicaltota}=\mathsf{scirawtota}/\mathsf{calrawtota}\cdot\mathsf{caltemplate}$
and the calibrated visibility by $V(\lambda)=\mathsf{scicalvis}=\mathsf{scirawvis}/\mathsf{calrawvis}=(\mathsf{scirawcor}/\mathsf{scirawtotg})\cdot(\mathsf{calrawtotg}/\mathsf{calrawcor})$.
Note that in this case $V(\lambda)\neq F_{\mathrm{cor}}(\lambda)/F_{\mathrm{tot}}(\lambda)$.

The entire data reduction and calibration was performed in the same
manner for all visibility points without any further individual adjustment
of the parameters, in order not to introduce any biases.

\subsection{$uv$ coverage\label{sub:circ_uv-coverage}}

\begin{figure*}
\centering
\includegraphics[%
  width=0.49\textwidth]{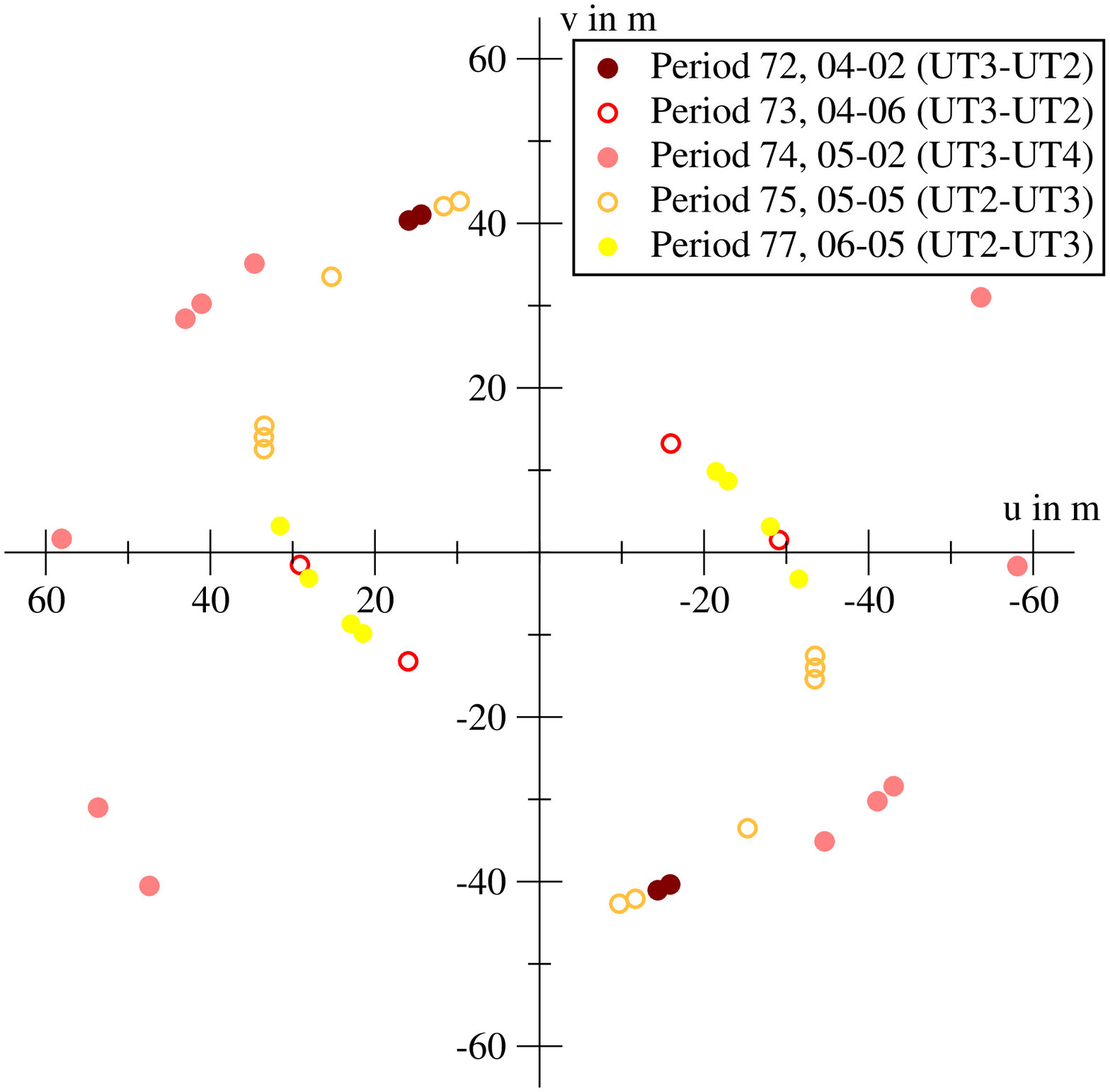}~~\includegraphics[%
  width=0.49\textwidth]{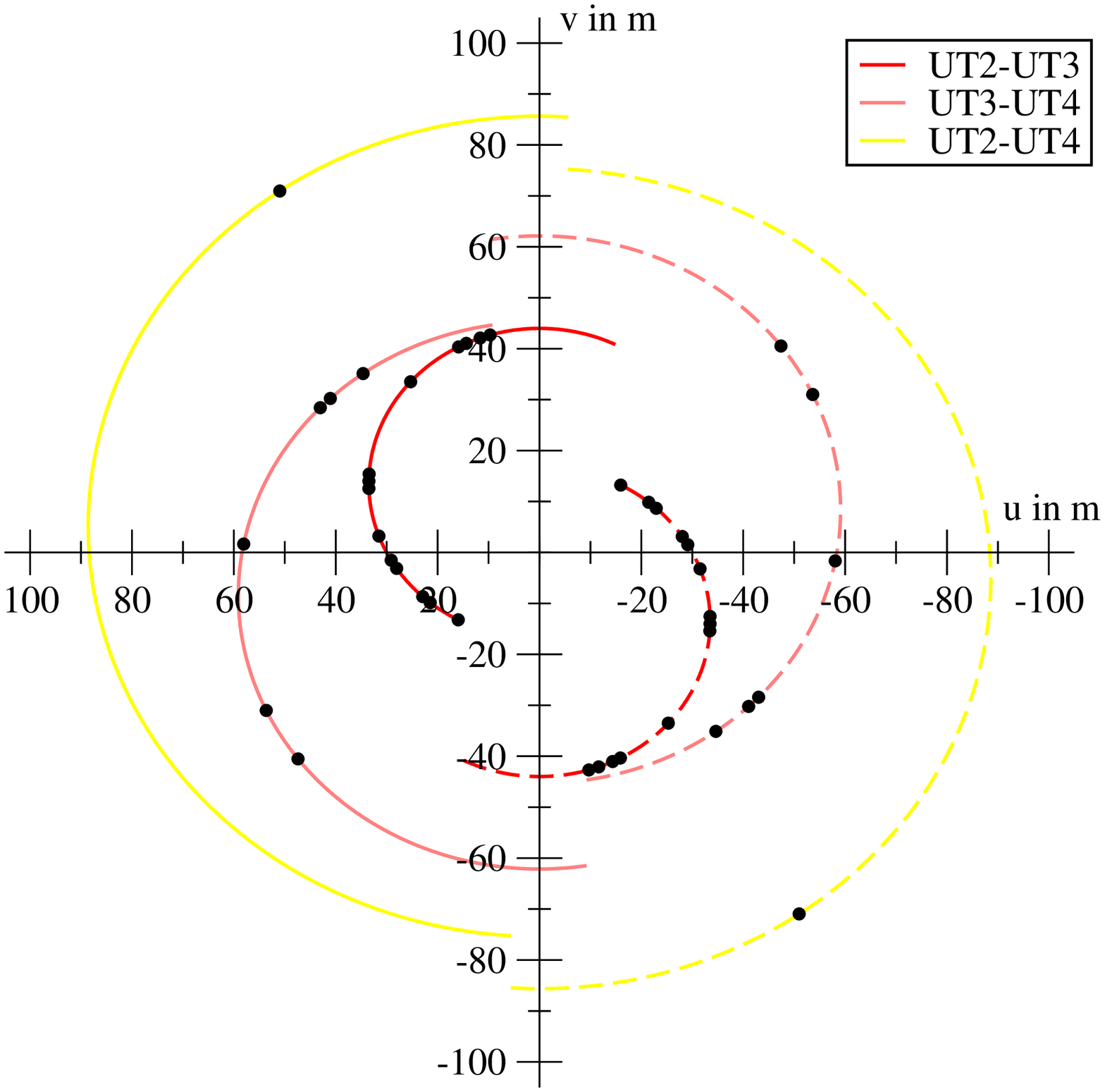}
\caption{Coverage of the $uv$ plane obtained for the Circinus galaxy with
MIDI, the MID-infrared Interferometric instrument at the VLTI. The
left panel shows the position of the observed fringe tracks in the
$uv$ plane for the two shorter baselines, UT2~--~UT3 and UT3~--~UT4.
The points are coded according to the epoch of observation. The right
panel shows all the 21 observed fringe tracks as well as the $uv$
tracks of the baselines for the three combinations of unit telescopes
that were used.}
\label{fig:circ_uv-coverage}
\end{figure*}

The interferometric data collected on Circinus constitute the most
extensive interferometric data in the infrared or at shorter wavelengths
of any extragalactic source to date. The coverage of the $uv$ plane
achieved by all of the MIDI observations is shown in Fig.~\ref{fig:circ_uv-coverage}.%
 Each point represents the position of a fringe track leading to a
visibility point. Because we measure the interferometric signal of
a real function (namely the brightness distribution) which has a symmetric
Fourier transform, there are two points for each measurement, symmetrical
to the centre. The different colours denote the different epochs of
observation. We observed 21 fringe tracks and 31 usable photometries
(note this is not the double of the fringe track number as it should
optimally be; there are only 15 usable A photometries and 16 usable
B photometries) for Circinus as well as 12 fringe tracks plus 24 photometries
for HD\,120404, the calibrator star. From these data, we were able
to reconstruct a total of 21 visibility points.

The distribution of the measurements in the $uv$ plane shown in Fig.~\ref{fig:circ_uv-coverage}
has concentrations at several locations, that is, two or three points
at almost the same location. This was caused by consecutive measurements
of the interferometric signal. The differences between such measurements
can be considered as indications for the accuracy of the measurements
themselves. Several measurements also share the same photometry or
calibrator data. In a strict sense, the true number of absolutely
independent measurements is thus only on the order of 12, which is
the number of independently measured calibrator data sets.

The observations were performed with only three baseline configurations:
UT2~--~UT3, UT2~--~UT4 and UT3~--~UT4. This causes the visibility
points to lie on arcs in the $uv$ plane, which correspond to the
classical $uv$ tracks known from radio interferometry. The $uv$
tracks for these three baseline configurations are plotted as continuous
lines in the right panel of Fig.~\ref{fig:circ_uv-coverage}. The
$uv$ coverage is far from optimal, as there are several regions lacking
measurements: at a position angle of $-20^\circ$, no measurements were
obtained at all and at a position angle of $+20^\circ$ only one baseline
length ($\mathit{BL}=40\,\mathrm{m}$) was observed. The void regions
therefore need to be filled in by future observations. In fact, the
current coverage is unsuitable to attempt any image reconstruction
as commonly done with radio interferometry data. There, a flux distribution
can be directly determined from the data by an inverse Fourier transform
and a prescription, such as the {}``clean algorithm''. We tried
applying such an algorithm to our data. However, the form of the PSF
(the {}``beam'') corresponding to our $uv$ coverage is very bad
and shows strong periodic patterns, thereby dominating the outcome
of the reconstruction process. The results of such an attempt are
poor and hard to interpret, reflecting only general properties of
the source. These general properties are much better constrained by
models (see Sect.~\ref{sec:circ_modelling}).

The differential phase (see Sect.~\ref{sub:circ_data-reduction})
is less than $20^\circ$ for almost all baselines. At the shortest baselines,
we find some evidence for a phase shift and there is also weak evidence
for phase shifts at positions where the visibilities show unexpected
behaviour (especially for sudden downturns at the edge of the spectal
range covered). Nevertheless, we find no evidence for a major asymmetry
in the source. We therefore postpone a detailed analysis of the differential
phase information and concentrate in this paper on the interpretation
of the visibility amplitude only.

\section{Results\label{sec:circ_results}}

\subsection{Total flux\label{sub:circ_total-flux}}

\begin{figure*}
\centering
\includegraphics{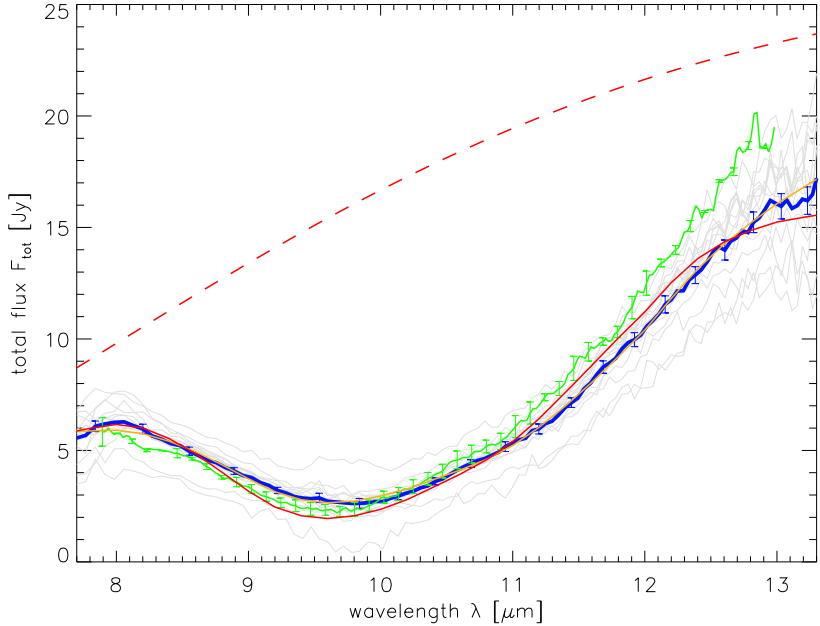}
\caption{Spectrum of the total flux of the Circinus nucleus in the N band.
The light grey curves are the 15 individual measurements of the total
flux; error bars were omitted for the sake of clarity. The thick blue
curve is the arithmetic mean of all the individual measurements, the
error bars indicating the reduced standard deviation. The spectrum
measured by \citet{1991Roche} is shown in green for comparison. The
yellow curve shows the fit of the physical model to the total flux
only (fit 1), while the red curve shows the fit to the full data set
consisting of the total flux and the correlated fluxes (fit 2). The
dashed red curve is the emission for the latter fit, without including
the extinction by dust. For details on the modelling, see Sect.~\ref{sub:circ_physical-model}.}
\label{fig:circ_total-flux}
\end{figure*}

Figure \ref{fig:circ_total-flux} shows the spectrum of the calibrated
total flux, $F_{\mathrm{tot}}(\lambda)$, of the Circinus nucleus.%
 In several of the individual measurements (light grey lines), enhanced
noisiness is apparent between $9.5$ and $10.0\,\mu\mathrm{m}$, which
is due to ozone absorption in the Earth's atmosphere. This is aggravated
by the fact that the ozone feature coincides with the minimum of the
flux in the intrinsic spectrum of Circinus. The individual measurements
are all consistent with the average (thick, blue line), when the associated
individual errors (not shown in Fig.~\ref{fig:circ_total-flux})
are considered. The curves with the largest deviations from the average
come from measurements in 2004, which also have larger individual
errors. The variations are mainly caused by imperfect background subtraction
for early data or observations at high airmasses. We clearly see an
increase in the data quality after the first measurements. The edges
of the N band are located at $8.2\,\mu\mathrm{m}$ and $13.0\,\mu\mathrm{m}$.
Towards shorter and longer wavelengths, the atmosphere has a high
opacity due to water absorption. Given the large number of measurements,
the averaged total flux is nevertheless usable beyond these boundaries. 

The broad absorption feature dominating most of the spectrum is due
to silicate absorption. We see no evidence for any line emission or
emission of Polycyclic Aromatic Hydrocarbons (PAH), as observed at
larger distances from the nucleus \citep{2006Roche} or in larger
apertures. The nuclear spectrum measured by \citeauthor{2006Roche}
agrees with our measurement of the total flux within $20\,\%$ (see
Fig. \ref{fig:circ_total-flux}). The differences are probably due
to the higher spectral resolution ($R\sim125$) and the different
orientation of the slit in the \citeauthor{2006Roche} data. The larger
aperture data from ISO \citep{1996Moorwood} and Spitzer (unpublished)
lie $\sim40\,\%$ above the MIDI total flux as they include the flux
from the circumnuclear starburst.

\subsection{Interferometric data\label{sub:circ_interferometric-data}}

\begin{figure*}
\centering
\includegraphics{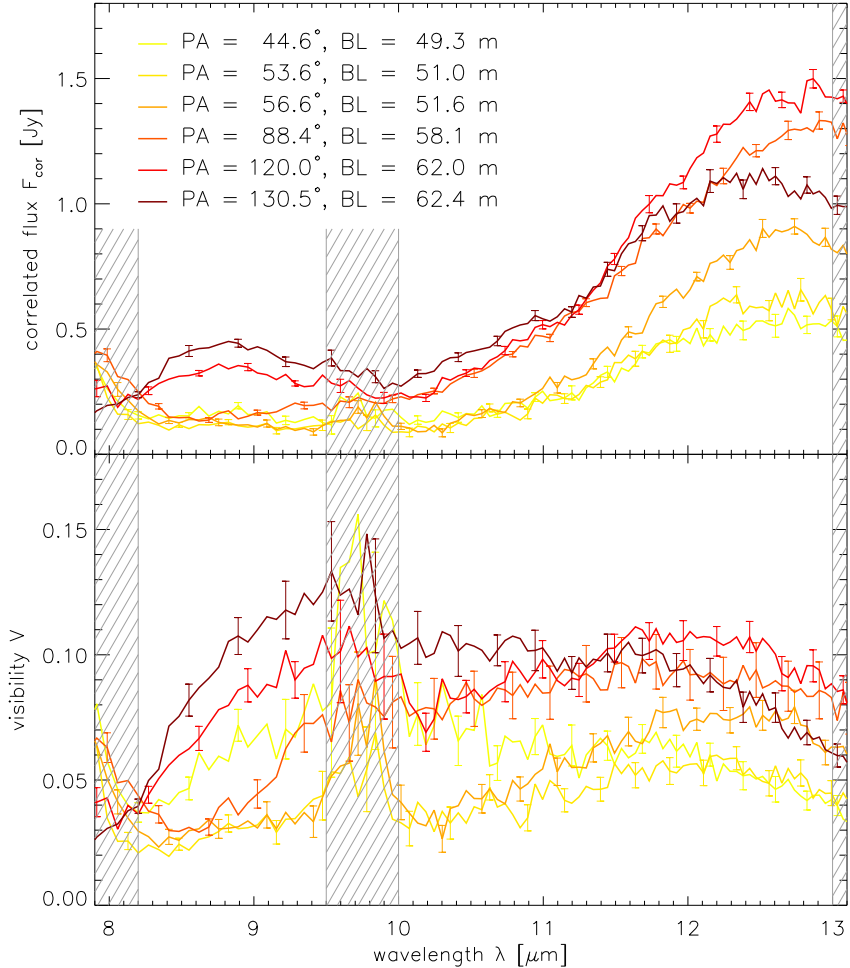}
\caption{Top panel: Spectra of the correlated fluxes, as observed during the
night of 2005 Mar 01. Bottom panel: Visibilities for the same observations.
During the course of these observations, the position angle ($\mathit{PA}$)
changed by $\sim90^\circ$, while the projected baseline length ($\mathit{BL}$)
stayed roughly the same. From the angular dependence of the correlated
fluxes, direct implications on the morphology can be derived: the
source observed is considerably more extended in the direction of
$\mathit{PA}\sim50^\circ$ than in the direction of $\mathit{PA}\sim130^\circ$.
Areas affected by atmospheric absorption are hatched in grey.}
\label{fig:circ_corvis}
\end{figure*}

For all baselines observed, the visibility is well below the point-source
value of $V=1$, indicating that the emission region is well resolved
with our interferometric resolution of $\lambda/2B\leq40\,\mathrm{mas}$.
As an example, we show in Fig.~\ref{fig:circ_corvis} the correlated
fluxes $F_{\mathrm{cor}}(\lambda)$ and visibilities $V(\lambda)$
obtained on 2005 Mar 01.%
 During this night, observations were performed using the baseline
UT3~--~UT4. The baseline length only increased from $50$ to $60\,\mathrm{m}$,
while the position angle underwent a major change from $44^\circ$ to $131^\circ$.
As in the total flux, the broad dip at $10\,\mu\mathrm{m}$ is due
to silicate absorption. The quality of the data at the edges of the
wavelength range and in a window between $9.5$ and $10.0\,\mu\mathrm{m}$
(hatched region) deteriorates significantly due to atmospheric water
and ozone absorption. We therefore only trust the spectra within these
boundaries, that is, from $8.2\,\mu\mathrm{m}$ to $9.5\,\mu\mathrm{m}$
and from $10.0\,\mu\mathrm{m}$ to $13.0\,\mu\mathrm{m}$.

A clear increase of the correlated flux and of the visibility can
be observed with increasing position angle. This increase is most
pronounced in the correlated fluxes longward of $11\,\mu\mathrm{m}$.
We directly interpret this as an angle-dependent change of size of
the emitting source: at position angles of $44^\circ$ to $57^\circ$ the correlated
flux and the visibility are low, \emph{i.e.} the source is more resolved
and hence the emission extended, while at angles of $120^\circ$ and $131^\circ$
the correlated flux and visibility are higher, \emph{i.e.} the source
is less resolved and consequently the emission must be less extended.
The angle-dependent size changes must be even stronger than Fig.~\ref{fig:circ_corvis}
insinuates, as there is a small increase in baseline length towards
larger position angles, that would lead to a decrease of the correlated
flux and visibility for a round and smooth source. This is a direct
and completely model independent evidence for an elongated dust structure.
This elongation is perpendicular to the ionisation cone and the outflow
at $\mathit{PA}=-40^\circ$ in the Circinus galaxy.

To analyse the emissivity and temperature structure of the target
either the visibility and the correlated flux can be used. Both approaches
have advantages and disadvantages. The visibility quantifies the degree
of {}``resolvedness'' of the source, that is, it contains geometrical
information on the source. The correlated flux reflects only the emission
from the unresolved part of the source, which per se does not contain
any geometrical information. It only receives a geometrical meaning
in comparison to the total flux or to correlated fluxes at different
baselines. From our observations we know that the photometry is the
most uncertain part of the measurement and we have fewer independent
measurements of photometries than fringe tracks. As described in Sect.~\ref{sub:circ_data-reduction},
the correlated fluxes are independent of errors in the photometry;
the visibility, however, always contains the photometry including
its uncertainties. For the analysis and the modelling, we consider
both the visibilities and the correlated fluxes as comparably useful
and we will use both for our modelling of the data. Furthermore, we
find that the essential inferences deduced from either the visibilities
or the correlated fluxes are the same; this is reassuring.

\section{Modelling\label{sec:circ_modelling}}

To interpret the visibilities in terms of a flux distribution and
to deduce physical properties of the source, the data need to be fitted
by models, because the $uv$ coverage is incomplete (see Sect.~\ref{sub:circ_uv-coverage})
and we lack absolute phase information. Our approach to model the
data consists of several steps of increasing complexity, minimising
any preconceptions about the source. The goal is to extract properties
of the mid-infrared emission, such as size, elongation, depth of the
silicate feature, etc., to sketch a picture of the nuclear dust distribution
in the Circinus galaxy.

The first step of modelling is aimed at estimating the characteristic
size of the emission region using a simple, one dimensional model.
From the discussion in Sect.~\ref{sub:circ_interferometric-data}
it is clear that a one dimensional model is too simple to explain
the MIR source in the nucleus of the Circinus galaxy correctly. More
complex, two dimensional models are necessary. Our next step in modelling
is hence a purely geometrically description of the source brightness
distribution for each wavelength bin. This second, two dimensional
model lead to the creation of a third, physically motivated model
which also attaches physical properties to the emission region. It
turns out that the latter model comprehends the main results of the
first two modelling steps. For this reason, only the physical model
will be presented here while the descriptions of the one-dimensional
and the purely geometrical model are found in App.~\ref{sub:circ_1d-model}
and App.~\ref{sub:circ_geo-model}. 

All of the modelling was performed using the Interactive Data Language
(IDL).

\subsection{Model selection\label{sub:circ_model-selection}}

For both the purely geometrical model as well as for the physical
model we explored a multitude of geometrical shapes in order to match
the data. The shapes include point sources, unifom disks, elliptical
disks, elliptical Gaussian distributions and combinations thereof.
We find that models including uniform disks produce results of similar
$\chi^{2}$ values as Gaussian distributions. The brightness distribution
of such models has sharp edges. The same is true for models including
point sources (\emph{e.g.} representing the hot inner edges of the
torus). Sharp edges have the great disadvantage that they produce
oscillations in the $uv$ plane: the $uv$ plane is not uniform, but
it has many minima and maxima in a quasi periodic pattern. Additionally,
phase shifts occur at the minima in the $uv$ plane. These are not
observed.

For this reason, we favour uniform models with no sharp edges to describe
our data. This restricts our conclusions to more general statements
about the properties of the mid-infrared source of Circinus. However,
we are confident that these properties are intrinsic to the source
and neither an artefact of the specific model nor of the distribution
of visibility points in the $uv$ plane. To be able to make clearer
statements about the detailed structure of the source, a more complete
coverage of the $uv$ plane is required and we are currently preparing
to fill in the most obvious holes in the distribution of the visibility
points, as far as this is possible with the baselines of the VLTI.

\subsection{2D physical model\label{sub:circ_physical-model}}

The physical model consists of two concentric black body Gaussian
emitters behind an extinction screen. For the absorption we used the
wavelength dependency of the extinction of interstellar dust from
\citet{2005Schartmann}, but modified the profile of the silicate
absorption feature in the wavelength range from $8.0$ to $12.7\,\mu\mathrm{m}$
using data from \citet{2004Kemper}. The latter determined the feature
profile towards the galactic centre and we thus expect their profile
to be our best guess for the line of sight towards the nucleus of
the Circinus galaxy. To obtain our template absorption profile $\tau(\lambda)$,
the entire optical depth curve was normalised to 1 at the maximum
absorption depth of the silicate profile: $\tau(9.7\,\mu\mathrm{m})=1$.

Our two component model has the following functional form:

\begin{equation}
\begin{array}{ccc}
F(\lambda,\alpha,\delta) & = & f_{1}\cdot G_{1}(\lambda,\alpha,\delta)\cdot\mathrm{e}^{-\tau_{1}\cdot\tau(\lambda)}\\
 & + & f_{2}\cdot G_{2}(\lambda,\alpha,\delta)\cdot\mathrm{e}^{-\tau_{2}\cdot\tau(\lambda)}\end{array}\label{eq:circ_phys-model-flux}\end{equation}
where $\tau_{1}$, $\tau_{2}$ are the individual optical depths of
the absorption screens. The scaling factors $f_{1}$ and $f_{2}$
are factors to scale the intensity of the black body Gaussian distributions
$G_{1}$ and $G_{2}$. The Gaussian distributions in turn are defined
by\begin{align}
G_{n}(\lambda,\alpha,\delta)=\exp\bigg[ & -4\ln2\cdot\left({\textstyle \frac{\alpha\cos\phi+\delta\sin\phi}{r_{n}\cdot\Delta_{n}}}\right)^{2}\\
 & -4\ln2\cdot\left({\textstyle \frac{\alpha\sin\phi-\delta\cos\phi}{\Delta_{n}}}\right)^{2}\bigg]\cdot F_{\mathrm{bb}}(T_{n},\lambda)\nonumber \end{align}
for each of the two components $n=1,2$. The parameters of the Gaussians
($r_{1}$, $r_{2}$ for the axis ratios, $\Delta_{1}$ and $\Delta_{2}$
for the FWHM and $\phi$ for the orientation) are the same as for
the purely geometrical model (see App.~\ref{sub:circ_geo-model}),
except that they are not wavelength dependent. $F_{\mathrm{bb}}$
is the black body intensity which depends on the temperature $T_{n}$.
This model has 11 free parameters: the FWHM $\Delta_{n}$, the axis
ratio $r_{n}$, the temperature $T_{n}$, the flux normalisation $f_{n}$
and the optical depth $\tau_{n}$ for each of the two components $n=1,2$.
The common position angle $\phi$ is the eleventh parameter%
\footnote{Decoupling $\phi$ for the two components leads the fits to converge
on a few dominant groups of visibility points only. This is the reason
for keeping a single orientation angle for both components.%
}. Using these parameters, the flux distributions at several wavelengths
were generated and Fourier transformed. From the resulting distribution
in the $uv$ plane, the visibilities corresponding to the observed
baselines, $V_{i}^{\mathrm{mod}}(\lambda)=\left|\mathcal{V_{\lambda}}(u_{i},v_{i})\right|$,
were extracted. Additionally, the total flux was determined by integrating
over the entire emission region: $F_{\mathrm{tot}}^{\mathrm{mod}}(\lambda)=\int\int F(\lambda,\alpha,\delta)\,\mathrm{d}\alpha\,\mathrm{d}\delta$.
The comparison to the observations was performed using both the correlated
fluxes and the total flux, because here the precise values should
match and not only the flux ratios. The correlated fluxes for the
model were obtained by multiplying the modelled visibilities with
the total flux of the model, \emph{i.e.} $F_{i}^{\mathrm{mod}}(\lambda)=V_{i}^{\mathrm{mod}}(\lambda)\cdot F_{\mathrm{tot}}^{\mathrm{mod}}(\lambda)$.
These could then be compared to the data. The model parameters were
optimised using the Levenberg-Marquardt least-squares minimisation.
The $\chi^{2}$ to be minimised was: \begin{equation}
\chi^{2}={\displaystyle \sum_{\lambda}\sum_{i=0}^{21}\left(\frac{F_{i}^{\mathrm{obs}}(\lambda)-F_{i}^{\mathrm{mod}}(\lambda)}{\sigma_{F_{i}^{\mathrm{obs}}}(\lambda)}\right)^{2}},\label{eq:circ_phys-model-chisq}\end{equation}
where the sum over $i$ is the summation over the 22 observed baseline
orientations (21 baselines plus the total flux as a baseline with
a baseline length of $\mathit{BL}=0\,\mathrm{m}$). The $\sigma_{F_{i}^{\mathrm{obs}}}(\lambda)$
are the errors of the correlated fluxes.

Two fits were performed: first the model was only optimised to fit
the total flux (in the following {}``fit 1''). In this case, several
parameters ($r_{1}$, $r_{2}$, $f_{1}$, $f_{2}$ and $\phi$) were
held fixed as they were degenerate. Then the model was fitted to the
total data set including both the total flux and the 21 correlated
fluxes ({}``fit 2''). In case of the fit to the total flux only,
there is hence no summation over $i$ in Eq.~\ref{eq:circ_phys-model-chisq},
as we simply had to use $i=0$. The result of the fitting process
is summarised in Table \ref{table:circ_phys-model-params}.%
\begin{table}

\caption{Best fit parameters for the two component black body Gaussian models.
The first column is for a fit of the total flux only (fit 1), the
second column for a fit to both the total flux and the correlated
fluxes (fit 2). The values in brackets were not fitted, but held fixed.}

\label{table:circ_phys-model-params}

\centering

\begin{tabular}{ccc}
\hline 
\hline parameter&
fit 1&
fit 2\tabularnewline
\hline
$\Delta_{1}$ {[}mas{]}&
23.6&
21.1\tabularnewline
$r_{1}$&
(0.25)&
0.21\tabularnewline
$\tau_{1}$&
1.17&
1.18\tabularnewline
$T_{1}$ {[}K{]}&
479.6&
333.7\tabularnewline
$f_{1}$&
(1.00)&
1.00\tabularnewline
$\Delta_{2}$ {[}mas{]}&
279.0&
96.7\tabularnewline
$r_{2}$&
(1.00)&
0.97\tabularnewline
$\tau_{2}$&
2.85&
2.22\tabularnewline
$T_{2}$ {[}K{]}&
144.4&
298.4\tabularnewline
$f_{2}$&
(1.00)&
0.20\tabularnewline
$\phi$ {[}°{]}&
(60.0)&
60.9\tabularnewline
$\chi^{2}/N_{free}$&
0.56&
36.86\tabularnewline
\hline
\end{tabular}
\end{table}

\begin{figure*}
\centering
\includegraphics{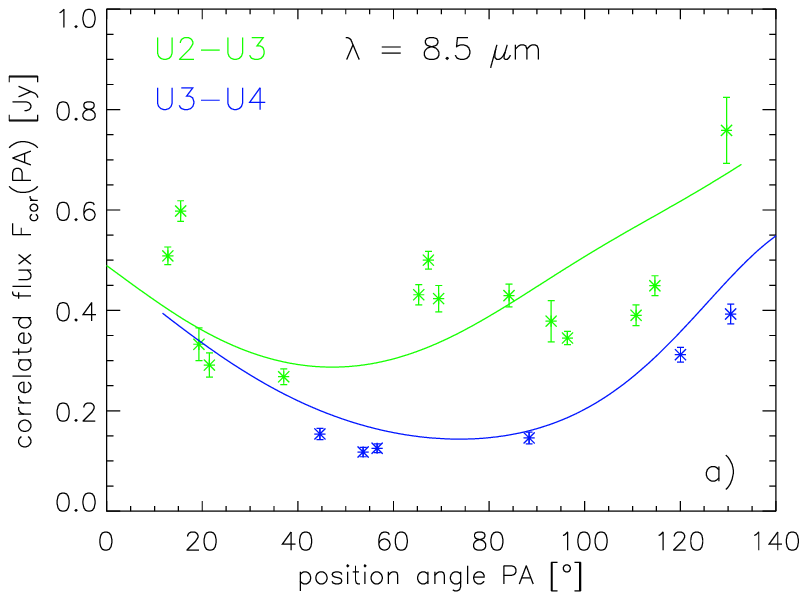}~~~\includegraphics{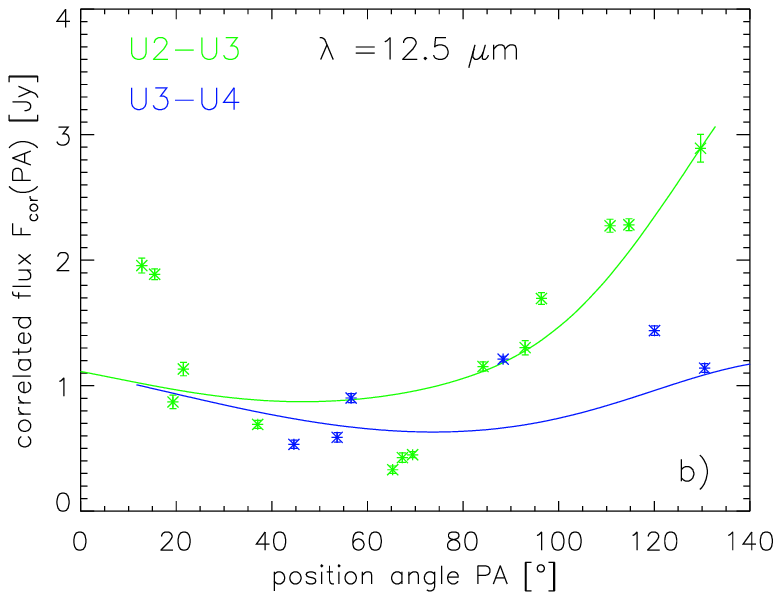}
\caption{Comparison of the correlated flux predicted by the physical model
(fit 2) with the observations for two distinct wavelengths: (a) $8.5$
and (b) $12.5\,\mu\mathrm{m}$. The measured fluxes are plotted with
asterisks; the continuous lines are the modelled fluxes. Note the
different flux ranges for the different wavelengths.}
\label{fig:circ_phys-model-compare}
\end{figure*}

A total of 21 wavelength bins were used for the fitting, so that fit
1 has $N_{\mathrm{free},1}=21\cdot1-11=10$ degrees of freedom, while
fit 2 has $N_{\mathrm{free},2}=21\cdot22-11=451$. The resulting total
fluxes for the two models are plotted in Fig.~\ref{fig:circ_total-flux}
in orange (fit 1) and red (fit 2) on top of the measured total flux.
The observed correlated fluxes are compared to fit 2 in Fig.~\ref{fig:circ_phys-model-compare}
at the two wavelengths already previously featured: $8.5$ and $12.5\,\mu\mathrm{m}$.%
 A comparison of fit 2 with all of the data is given in App.~\ref{sec:appendix}.

\begin{figure}
\centering
\includegraphics[scale=0.6]{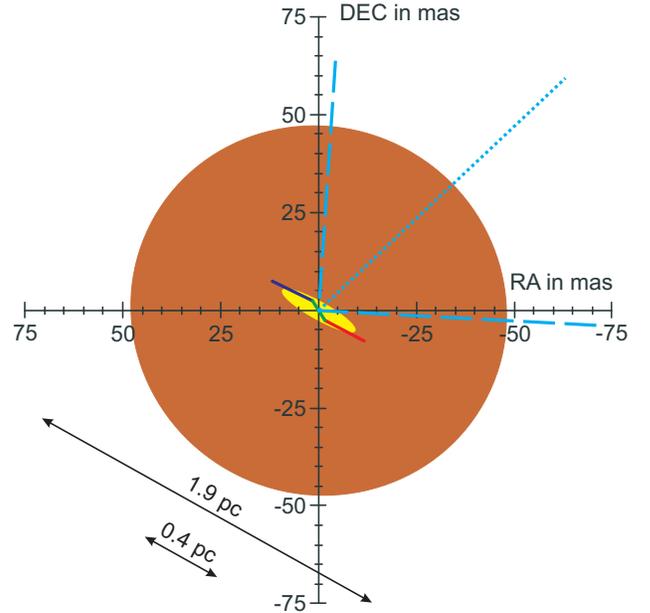}
\caption{Sketch of the physical model (fit 2): a highly elongated warm emission
region with a temperature of $\sim330\,\mathrm{K}$ (yellow) is surrounded
by an extended, almost round and slightly cooler emission region with
a temperature of $\sim300\,\mathrm{K}$ (brown). In the centre of
the sketch, the location of the $\mathrm{H}_{2}\mathrm{O}$ maser
emitters in a disk from \citet{2003Greenhill} is overplotted: the
blue line to the north-east traces the receeding masers and the red
line to the south-west the approaching masers. The dashed light blue
line traces the edge of the observed ionisation cone, the dotted line
is the cone axis.}
\label{fig:circ_phys-model-sketch}
\end{figure}

\begin{figure}
\centering
\includegraphics{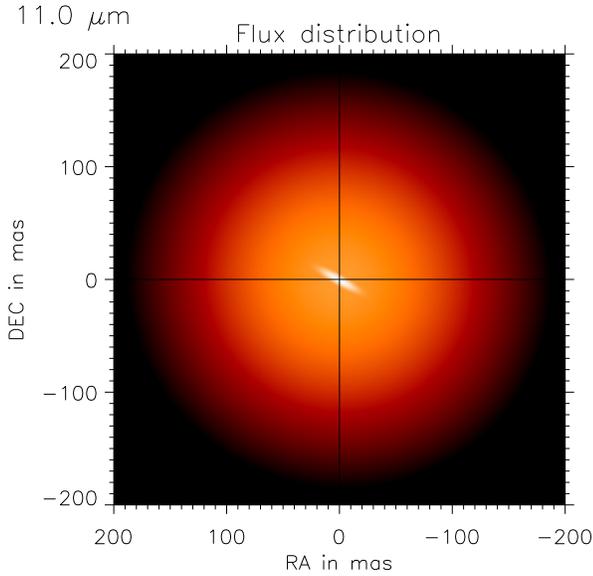}
\caption{Flux distribution of the physical model (fit 2) at $11\,\mu\mathrm{m}$
in logarithmic scaling.}
\label{fig:circ_phys-model-flux}
\end{figure}

The two components of this model have similar properties to those
of the purely geometrical model (App.~\ref{sub:circ_geo-model}):
there is a smaller elongated component and a larger, nearly round
component. A sketch of the result of fit 2 is depicted in Fig.~\ref{fig:circ_phys-model-sketch}%
 and the flux distribution at $11\,\mu\mathrm{m}$ can be seen in
Fig.~\ref{fig:circ_phys-model-flux}.%
 The smaller component has a FWHM of $\Delta_{1}=21\,\mathrm{mas}$,
which corresponds to $0.4\,\mathrm{pc}$ at the distance of Circinus.
It is highly flattened ($r_{1}=0.21$) and has a temperature of $T_{1}=334\,\mathrm{K}$.
The distribution is found to be an optically thick black body ($f_{1}=1$).
The second component is significantly larger, as a large part of the
observed flux ($\sim90$\%) is resolved with our interferometric set-up.
It has a FWHM of $\Delta_{2}=97\,\mathrm{mas}$, which corresponds
to $2.0\,\mathrm{pc}$, only a very small ellipticity and a temperature
of less than $300\,\mathrm{K}$. The emissivity of the black body
radiation from this component is scaled by a factor of $f_{2}=0.20$.
This low scaling factor is driven by the rather extended flux distribution,
which is necessary to explain the low visibilities, and the requirement
not to overpredict the measured total flux. Note that the scaling
factor is not to be interpreted as a geometrical covering factor.
Together with the charactersitic temperature $T_{2}$, it rather is
a formal description of a significantly more complex multi-temperature
system. This is underlined by the fact that there is a degeneracy
between the temperature and the scaling factor in the fit. When fitting
only the total flux, we see a larger temperature difference between
the two components than for the full fit. In both cases, fit 1 and
2, the temperatures are the least constrained parameters. We therefore
consider the $330\,\mathrm{K}$ for the inner component as a lower
limit for the highest dust temperature and the $300\,\mathrm{K}$
as an upper limit for the cool component. Our data rules out any significant
contribution from a truly hot component with temperatures $T>1000\,\mathrm{K}$,
\emph{i.e.} close to the sublimation temperature of the dust. This
is consistent with \citet{2004Prieto}, who also found no evidence
for hot dust. The conclusion drawn from the growth of the size of
the emitter with wavelength (Sect.~\ref{sub:circ_1d-model}), namely
that the temperature of the emitter decreases with increasing distance
to the nucleus, is confirmed by the lower temperature of our larger
component. The small component only contributes a minor fraction to
the total flux. This can be seen in the top left graph of Fig.~\ref{fig:circ_phys-model-compcor}:
the flux contribution by the extended component is delineated by a
dashed line and that by the compact disk component by a dotted line.
In contrast to the observations of NGC~1068 \citep{2004Jaffe1,2006Poncelet},
the correlated fluxes in the Circinus galaxy do not show a relative
increase of the flux to shorter wavelengths with respect to the flux
at longer wavelengths. This shows that there is only a moderate increase
of dust temperature towards the centre.

The model used here is a parametrised global model and does not match
all aspects of the data. The reduced $\chi^{2}$ of $37$ indicates
that deviations well above the 5$\sigma$-level are commonly found.
The aim of our model was to extract the overall properties of the
emitting source. The general behaviour of the correlated fluxes for
the different baseline orientations can be well reproduced, but not
the details. Looking at Fig.~\ref{fig:circ_phys-model-compare},
the model fluxes only show a single dip in the range of position angles
traced, due to the symmetry of the object when rotated by $180^\circ$.
The measurements exhibit considerably more rapid angular variations,
rejecting a smooth axisymmetric model. The dispersed fluxes (Fig.~\ref{fig:circ_phys-model-compcor})
disagree most strongly at the edges of the spectrum, at $8\,\mu\mathrm{m}$,
where a downturn in several of our correlated fluxes is seen, or at
$13\,\mu\mathrm{m}$, where the model deviates in both directions
with respect to the data. As there are also small changes in the differential
phase in these regions, we expect a further substructure of the source
to be responsible for this unusual behaviour.

\begin{figure}
\centering
\includegraphics{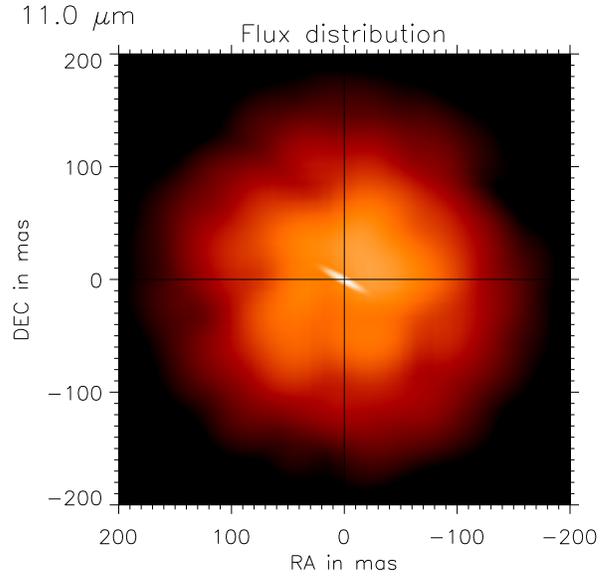}
\caption{Flux distribution of the physical model with added clumpiness for
the extended component at $11\,\mu\mathrm{m}$ in logarithmic scaling.}
\label{fig:circ_phys-clump-flux}
\end{figure}

\begin{figure*}
\centering
\includegraphics{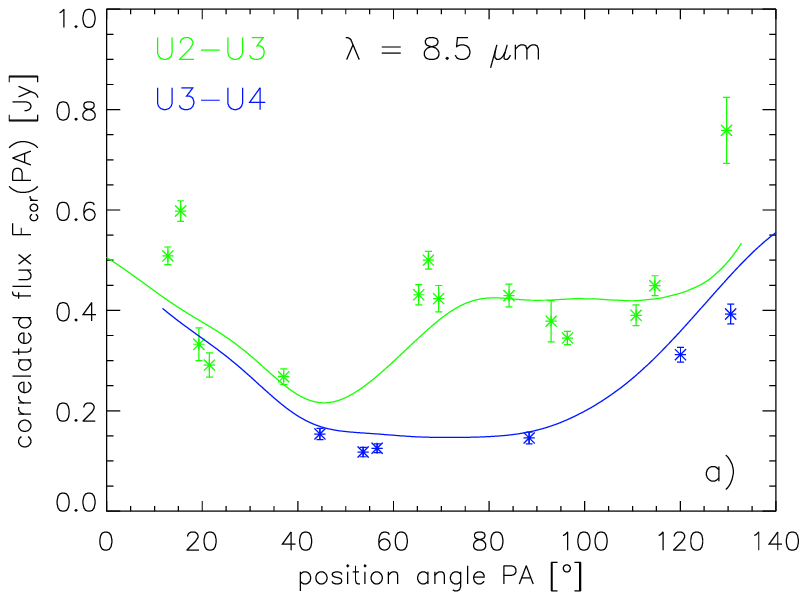}~~~\includegraphics{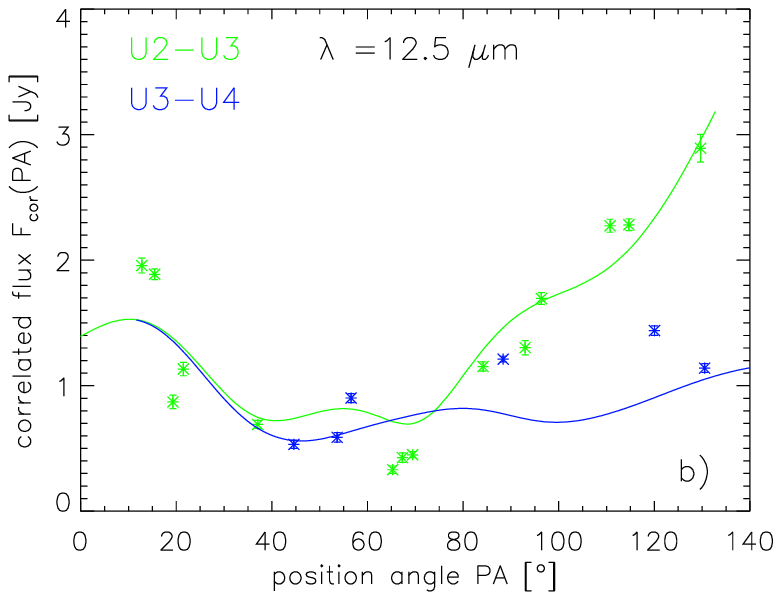}
\caption{Correlated flux of the physical model with simulated clumpiness for
the extended component: (a) $8.5$ and (b) $12.5\,\mu\mathrm{m}$.
The measured fluxes are plotted with asterisks; the continuous lines
are the modelled fluxes. The clumpiness introduces significant changes
in the correlated fluxes compared to the smooth model (see Fig.~\ref{fig:circ_phys-model-compare}).
From among 1000 clumpy models we found at least one (shown here),
which matches the data well. This shows that clumpiness of a certain
geometry can greatly improve the fit to the data.}
\label{fig:circ_phys-clump-compare}
\end{figure*}

To verify this possibility, we undertook an additional modelling experiment:
we tried to reproduce the low scaling factor of the larger component
through clumpiness: instead of multiplying $F_{1}$ with the factor
$f_{2}$, an uneven surface $C(\alpha,\delta)$ was used. $C(\alpha,\delta)$
can be interpreted as a screen simulating clumpiness. To make the
calculation consistent, the average value of $C$ has to reproduce
the scaling factor: $\left\langle C(\alpha,\delta)\right\rangle =f_{2}$
. The resulting flux distribution for such a clumpy model is shown
in Fig.~\ref{fig:circ_phys-clump-flux}.%
 In such a simple experiment, it is possible to find clump distributions
which trace the correlated flux much better (see for example Fig.~\ref{fig:circ_phys-clump-compare}).%
 The reduced $\chi^{2}$ drops to values of $\chi^{2}/N_{free}<25$.
This supports the assumption that such deviations are generated by
a substructure of the brightness distribution.

\section{Discussion\label{sec:circ_discussion}}

\begin{figure}
\centering
\includegraphics[scale=0.6]{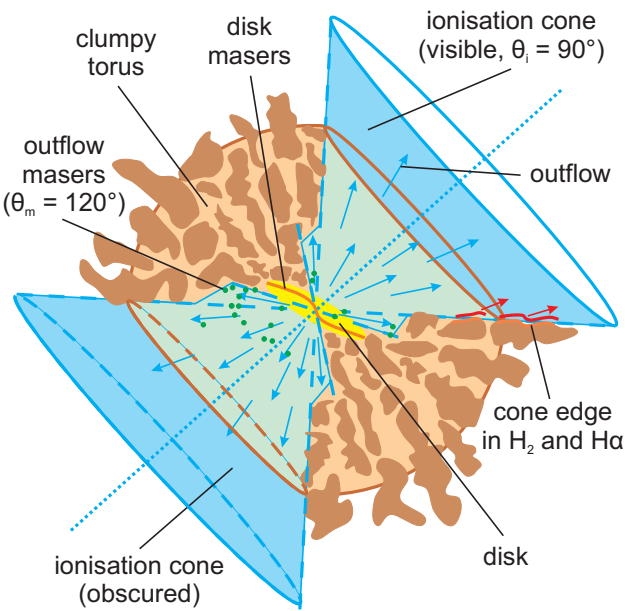}
\caption{Cut through the dusty torus in the central parsec-sized region of
the Circinus galaxy derived from the interferometric observations
in the MIR. See text for details.}
\label{fig:circ_discussion-torus-sketch}
\end{figure}

The modelling of our interferometric observations reveals the presence
of two components of dust emission at the very centre of the Circinus
galaxy: an extended ($r_{1}=1.0\,\mathrm{pc}$), warm ($T_{2}\sim300\,\mathrm{K}$)
and fairly round emission region showing strong silicate absorption
and a much smaller ($r_{2}=0.2\,\mathrm{pc}$), but only slightly
warmer ($T_{1}\sim330\,\mathrm{K}$) component that is highly flattened
and shows a less pronounced silicate absorption feature. We interpret
these two components as signs for a geometrically thick {}``torus''
of warm dust surrounding a warmer, disk-like structure (see Fig.~\ref{fig:circ_discussion-torus-sketch}).%
 The smaller and denser disk is seen at a high inclination and it
partly exhibits the silicate band in emission. It is surrounded by
the larger, less dense and most likely clumpy torus component which
gives rise to strong silicate absorption. The silicate absorption
profiles towards both the disk and the torus component are consistent
with the interstellar absorption profile observed towards the nucleus
of our own galaxy. This suggests that the dust composition is the
same. The result is interesting compared to that found for NGC~1068,
where the dust absorption towards the nucleus clearly does not agree
with the standard intergalactic absorption profile \citep{2004Jaffe1,2006Poncelet}.

To understand the structure of the nuclear dust distribution in AGN,
we developed hydrodynamical models of dusty tori (for a detailed description
see \citealt{2007MarcThesis}). These models indeed also show a disk-like
structure in the inner region which is surrounded by a geometrically
thick {}``torus''. The torus is not continuous but has a filamentary
structure (rather than clumpy) and the disk is turbulent.

In the following, we will develop a picture of the central parsec
region in Circinus which aims to explain the results of the MIDI measurements
in the context of the hydrodynamical model. In addition, the picture
needs to be consistent with the previously known properties of this
galactic nucleus.

\subsection{Extinction\label{sub:circ_extinction}}

To interpret the nuclear emission from Circinus and the silicate absorption
in the torus, we need to estimate the foreground extinction. Radiation
from the nuclear region of Circinus suffers extinction from two main
foreground absorbers: our own galaxy (Circinus is located at a galactic
latitude of $b=-4^\circ$) and the intrinsic absorption in the galactic
disk of the Circinus galaxy.

By observing the reddening of several stars in the vicinity of Circinus,
\citet{1977Freeman} estimate the visual extinction due to our own
galaxy to be only $A_{V}=1.50\pm0.15\,\mathrm{mag}$, as Circinus
lies in a window of low galactic extinction. The extinction towards
the nuclear region within Circinus itself is much less constrained
and estimates from near-infrared colors vary from $A_{V}=6\,\mathrm{mag}$
for a foreground screen to $A_{V}\gtrsim20\,\mathrm{mag}$ for a mix
of dust and stars \citep{2004Prieto,2006MuellerSanchez}. In the latter
case, the colours saturate so that higher extinctions cannot be ruled
out. The near-infrared observations which underlie this estimate probe
a region on scales of less than $10\,\mathrm{pc}$ distance from the
nucleus. Given the presence of a nuclear starburst \citep{2006MuellerSanchez},
a mix of dust and stars appears to be more realistic and we adopt
a foreground extinction, between us and $\sim2\,\mathrm{pc}$ from
the nucleus, of $A_{V}\sim20\,\mathrm{mag}$.

From the optical depth of the silicate feature seen in the extended
torus component, we can estimate the extinction towards those regions
of the torus where the warm dust emission arises. Our value of $\tau_{2}\sim2.2$
for the optical depth is at the lower end of the values found by \citet{2006Roche},
who cite a range of $2.2\leq\tau_{\mathrm{sil}}\leq3.5$, depending
on where they extract the spectrum along their slit. Given the low
spectral resolution of MIDI the minimum of the absorption trough may
be washed out, leading to lower optical depths. We therefore consider
the two values to be in good agreement, although our observations
rule out $\tau_{\mathrm{sil}}>3.0$ for the nucleus. In the following,
we will assume $\tau_{\mathrm{sil}}=2.5$ towards the larger ($r=1\,\mathrm{pc}$)
torus component.

We use the opacity scaling of \citet{2005Schartmann}, to convert
the opacity measured in the $9.5\,\mu\mathrm{m}$ silicate feature
to an equivalent visual extinction, $\tau_{V}=12.2\cdot\tau_{\mathrm{sil}}$.
Hence, the absorption in the visual is $A_{V}=1.09\cdot\tau_{V}\,\mathrm{mag}=13.3\cdot\tau_{\mathrm{sil}}\,\mathrm{mag}$.
For $\tau_{\mathrm{sil}}=2.5$, we obtain $A_{V}=33\,\mathrm{mag}$,
which means that there are an additional $11\,\mathrm{mag}$ of extinction
towards the torus region compared to the limit derived from the near-infrared
colours. Most likely, the extinction towards the accretion disk itself
is even higher.

In contrast to the MIDI observations of NGC~1068 \citep{2004Jaffe1,2006Poncelet},
we do not observe a further increase of the silicate absorption feature
when zooming into the nucleus with our interferometric observations.
Instead, the absorption feature appears to be less pronounced in the
correlated fluxes, which can also be inferred directly from the convex
shape of the visibilities (\emph{e.g.} from Fig.~\ref{fig:circ_corvis}).
In general, the silicate feature is expected to appear in emission
for lines of sight, where the (appropriately high) temperature decreases
away from the observer. This is the case for the directly visible
inner rim of the torus and, therefore, expected and observed for Seyfert
1 objects. In the opposite case (cold dust along the line of sight),
the silicate feature will appear in absorption, as seen in Seyfert
2 objects. A decrease of the silicate feature in absorption therefore
means that more and more of the Seyfert 1 type silicate emission is
able to shine through the torus body. These inferences are consistent
with a hidden Seyfert 1 nucleus in Circinus, a fact also suggested
by the observation of broad emission lines in polarised light \citep{1998Oliva}.

In summary, we conclude that the obscuration on our line of sight
towards the inner parsec of Circinus is composed of the following
three extinction components: extinction by our Milky Way, $A_{V}(\mathrm{MW})=1.5\,\mathrm{mag}$,
by the foreground dust within Circinus, $A_{V}(\mathrm{disk})=20\,\mathrm{mag,}$
and by the nuclear dust, $A_{V}(\mathrm{nuclear})=11\,\mathrm{mag}$.

\subsection{Energy budget\label{sub:circ_energy-budget}}

The energy emitted by the source in the N band can be estimated by
integrating the flux in the dereddened spectrum, giving a luminosity
in the N band of $L_{7.5-13.5\,\mu\mathrm{m}}=1.3\cdot10^{9}\,\mathrm{L}_{\odot}=5.1\cdot10^{35}\,\mathrm{W}$.

To estimate the total luminosity of the dust emission, we integrate
under the flux of our physical model (fit 2) from $\nu=0$ to $\infty$
without applying any absorption, that is $\tau_{1}=\tau_{2}=0$. Orientation
effects only affect significantly the smaller disk component. A correction
by assuming oblate spheroids seen edge-on leads to an increase of
the flux by about $20\,\%$. With this method, we obtain a total luminosity
from the dust of $L_{\mathrm{dust}}=4.1\cdot10^{9}\,\mathrm{L}_{\odot}$.
This is four times more than the luminosity measured in the N band
alone. Considering that we have not traced the contribution by dust
at $T\lesssim150\,\mathrm{K}$, we round our value up to $L_{\mathrm{dust}}\sim5\cdot10^{9}\,\mathrm{L}_{\odot}$
as our best estimate for the luminosity of the nuclear dust at $r<2\,\mathrm{pc}$
in the infrared. 

We can take this value to estimate the luminosity of the accretion
disk, $L_{\mathrm{acc}}$, which presumably is responsible for heating
the dust. In doing so, we have to take into account that not all of
the accretion luminosity is available for heating the torus. A significant
fraction of $L_{\mathrm{acc}}$ will escape along the directions of
the ionisation cones. With an opening angle of $90^\circ$ \citep{1997Veilleux,2000Wilson},
the total solid angle covered by the torus, as seen from the centre,
is $2\sqrt{2}\pi$. This corresponds to about 0.7 of the full sphere.
Assuming a radiation characteristic for an optically thick accretion
disk which is proportional to $\cos\theta$, about half of the radiation
is emitted in direction of the dust in the torus. That is, only half
of the energy from the accretion disk will be absorbed by the dust
and re-radiated in the infrared. If we thus assume $L_{\mathrm{dust}}\sim0.5\cdot L_{\mathrm{acc}}$,
this leads to a luminosity of the central energy source of $L_{\mathrm{acc}}\sim10^{10}\,\mathrm{L}_{\odot}$.

Consistent with our estimate, \citet{1996Moorwood} also report the
luminosity of the nucleus to be $\sim10^{10}\,\mathrm{L}_{\odot}$.
Their integration under the UV bump (the unobscured radiation) yields
the same luminosity as we observed in the MIR: $L_{\mathrm{UV}}=5\cdot10^{9}\,\mathrm{L}_{\odot}$.
\citet{1999Oliva} cite a total ionising radiation of the nucleus
on the order of $L_{\mathrm{ion}}=1-4\cdot10^{10}\mathrm{L}_{\odot}\cdot\sin^{-2}i$,
where $i$ is the inclination angle of the ionisation cone with respect
to our line of sight. For $i>60$ \citep{1998Elmouttie}, this translates
into $1-7\cdot10^{10}\,\mathrm{L}_{\odot}$. Luminosities of the AGN
in Circinus significantly higher than $10^{10}\,\mathrm{L}_{\odot}$
can be ruled out, as this would imply a higher MIR flux than observed.
The currently derived intrinsic $2$ to $10\,\mathrm{keV}$ luminosities
of the nuclear X-ray source are all below $5\cdot10^{8}\,\mathrm{L}_{\odot}$,
which is well below the energy needed to heat the dust. The dust is
mainly heated by soft X-rays and UV radiation rather than by radiation
in the $2$ to $10\,\mathrm{keV}$ band alone.

Finally, we can compare our luminosity estimate for the accretion
disk to the Eddington luminosity. For a black hole with a mass of
$M_{\mathrm{BH}}=1.7\cdot10^{6}\,\mathrm{M}_{\odot}$ \citep{2003Greenhill},
this luminosity is $L_{\mathrm{E}}=5.6\cdot10^{10}\,\mathrm{L}_{\odot}$.
With our value of $L_{\mathrm{acc}}\sim10^{10}\,\mathrm{L}_{\odot}$,
we find Circinus to accrete at $18\,\%$ of its Eddington luminosity.

\subsection{Temperature dependency and clumpiness\label{sub:circ_temperature-dependence}}

Less than a parsec from this central energy source, the dust is most
likely heated by the radiation from the central accretion disk alone
and not by a homogeneous energy input, such as by a starburst. The
discussion of the energy budget of the nucleus has shown that no further
energy source is necessary. A heating by a nuclear energy source is
also suggested by the decrease of the dust temperature to the outside.

\begin{figure*}
\centering
\includegraphics{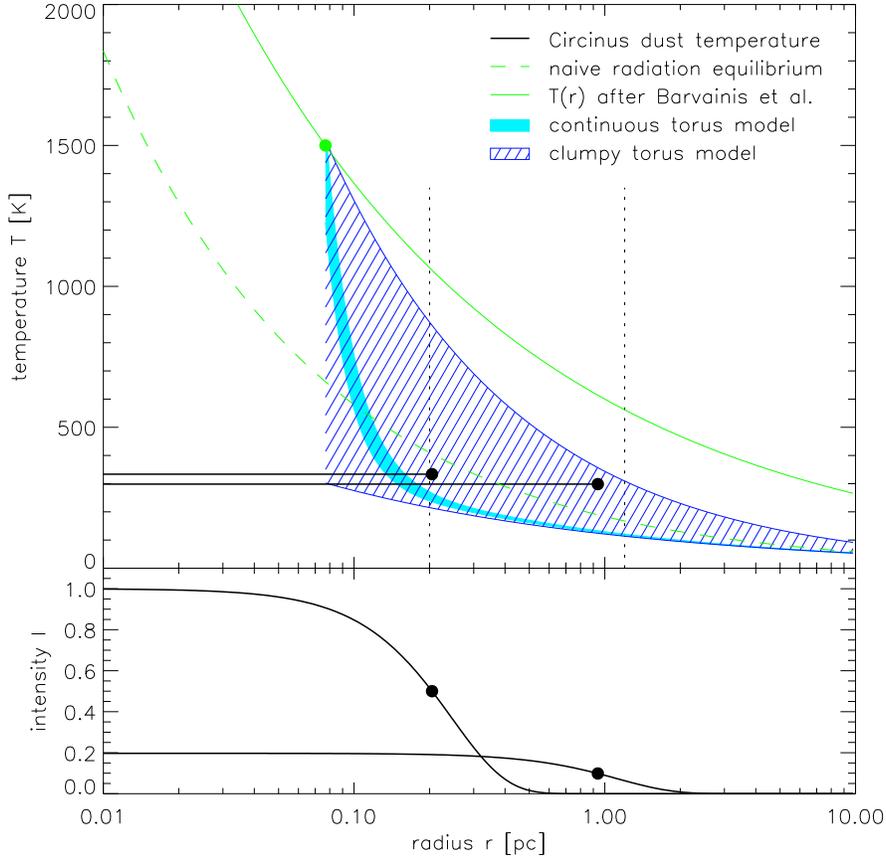}
\caption{Radial temperature gradients in the central dust distribution. The
radial temperature distribution of our simple {}``physical'' model
(thick black continuous lines) is compared to the temperature distributions
for a clumpy (hatched area) and continuous torus (light blue / grey
filled area) as well as to the temperature dependency of dust directly
illuminated by the central energy source (green / grey lines). The
point where the sublimation temperature of the dust, $T_{B}(r)=1500\,\mathrm{K}$,
is reached is marked by a green / grey circle. The lower part of the
graph shows the intensity of the Gaussian components (depending on
the scaling factor $f$). The HWHM of the Gaussian components are
marked by circles. The scales probed by our interferometric set-up
lie within the dotted vertical lines.}
\label{fig:circ_temperature-dependence}
\end{figure*}

The temperature decrease we observe is, however, very shallow and
the temperature of our large component is relatively high considering
the distance of its dust to the heating source: the dust has a temperature
of $\sim300\,\mathrm{K}$ out to distances of $1\,\mathrm{pc}$ from
the nucleus. To investigate the implications of this finding on the
properties of the dust distribution, we compare the dust temperature
in Circinus to the temperature expected for a direct illumination
of the dust and to the temperature inferred from our radiative transfer
modelling of continuous and clumpy AGN tori. The various temperature
dependencies are shown in Fig.~\ref{fig:circ_temperature-dependence}
and we will discuss them one by one in the following. %

Assuming a naive radiation equilibrium, the temperature of a dust
particle at the distance $r$ from the illuminating source with the
luminosity $L_{\mathrm{acc}}$ is given by\begin{equation}
T_{\mathrm{N}}(r)=194\left(\frac{L_{\mathrm{acc}}}{r^{2}}\frac{\mathrm{pc}^{2}}{10^{10}\,\mathrm{L_{\odot}}}\right)^{1/4}\,\mathrm{K}.\end{equation}
Taking into account the absorption and emission characteristics of
the dust grains leads to a more realistic description of the temperature
dependency \citep{1987Barvainis}. For emission in the infrared, the
temperature dependency then reads:\begin{equation}
T_{\mathrm{B}}(r)=624\left(\frac{L_{\mathrm{acc}}}{r^{2}}\frac{\mathrm{pc}^{2}}{10^{10}\,\mathrm{L_{\odot}}}\right)^{1/5.6}\,\mathrm{K}.\end{equation}
The two temperature dependencies $T_{\mathrm{N}}(r)$ and $T_{\mathrm{B}}(r)$,
calculated for $L_{\mathrm{acc}}=10^{10}\,\mathrm{L}_{\odot}$, are
plotted Fig.~\ref{fig:circ_temperature-dependence} by a green dashed
line and a green continuous line respectively (in the printed version
the green lines appear grey). The realistic case leads to significantly
higher temperatures than the naive case, as the individual dust particles
reradiate the energy less effectively than a black body. Here, we
have not taken into account the anisotropic radiation characteristic
of the central energy source, the accretion disk, which emits less
energy in direction of the dust distribution than in direction of
the ionisation cones (see Sect.~\ref{sub:circ_energy-budget}). This
leads to a lower temperature of the dust near the plane of the disk,
so that $T_{B}(r)$ should be considered as an upper limit for the
dust temperature at a certain radius. 

In the case of a continuous dust distribution, the temperature is
expected to drop steeply outside the inner funnel of the torus, where
the dust is directly illuminated by the central source. The dust at
larger radii is shielded from the nuclear heating and by consequence
at lower temperatures. For a clumpy torus, the temperature dependency
should be somewhat intermediate between the continuous case and the
radiation equilibrium for totally unobscured lines of sight, that
is, $T_{\mathrm{B}}(r)$. The behaviour is reproduced by our radiative
transfer modelling of AGN tori: the shaded areas show the temperature
ranges (from mean to maximum temperature) of the individual cells
in a continuous and a clumpy torus model (for a detailed description
of the modelling we refer to \citealp{2007Schartmann}). For this
purpose, the radii were rescaled so that the inner radii of the torus,
where the sublimation temperature is reached, coincide with the sublimation
radius predicted by $T_{B}(r)$. As expected, the continuous model
(filled, light blue / grey area) has a very strong drop in the temperature
behind the inner edge of the dust distribution, while the clumpy model
(dark blue, hatched area) has a much wider temperature range, including
lower temperature clouds at small distances from the nucleus and higher
temperatures at larger distances.

Finally, the thick black continuous lines trace the temperatures of
the two dust components in Circinus. The lines are drawn out to the
half width at half maximum (HWHM) of the respective component. From
the figure it is obvious that the temperatures seen by MIDI disagree
with a continuous dust distribution as the temperatures lie significantly
above the distribution and no sharp temperature rise to the centre
is seen. The high temperatures at large distances from the nucleus
can only be achieved when clouds also at larger distances have free
lines of sight to the heating source. As the temperatures lie within
the temperature range for a clumpy dust distribution, we conclude
that the dust most likely is distributed in a clumpy medium. All our
temperatures lie below those of $T_{B}(r)$, as expected for partial
shadowing of the clouds.

There are several further indications for a {}``clumpy'' or patchy
dust distribution in our data:

The low emissivity filling factor of the torus-like component in our
physical model is required by the data and best explained by an inhomogeneity
of the emitting dust. Patchy emission appears the only way for thermal
emission of dust at $T\sim300\:\mathrm{K}$ to be as extended as observed
here, while at the same time being consistent with the total flux
of the source.

The most compelling evidence, however, comes from the behaviour of
the correlated fluxes (a) with wavelength or (b) with changing baseline
(for the sake of simplicity we will only refer to the correlated fluxes
$F_{\mathrm{cor}}$ in this paragraph, nevertheless the same statements
are true for the behaviour of the visibilities $V$). The behaviour
observed cannot be explained by a smooth brightness distribution.

When considering the set of all visibility points observed at a single
wavelength, \emph{i.e.} $F_{\mathrm{cor},\lambda}(\mathit{PA})$,
the need for clumpiness becomes obvious when comparing Fig.~\ref{fig:circ_phys-model-compare}
to Fig.~\ref{fig:circ_phys-clump-compare}, as was already discussed
in Sect.~\ref{sub:circ_physical-model}. We have also compared the
variations we observe in our data to those that can be seen in the
correlated fluxes of our clumpy radiative transfer models \citep{2007Schartmann}:
when we calculate the correlated fluxes of such a torus model for
different position angles, $F_{\mathrm{cor},\lambda}(\mathit{PA})$,
we obtain variations of up to $50\,\%$ of the flux, similar to those
shown in Fig.~\ref{fig:circ_phys-clump-compare}. The variations
become stronger for longer baselines, where substructures are better
resolved.

Similarly, the need for a clumpy structure becomes obvious when considering
the individual visibility points at different wavelengths, that is,
the dependence of the correlated flux over the N band, $F_{\mathrm{cor},\mathit{PA}}(\lambda)$.
In this case, we observe wiggles in the correlated fluxes that can
only be reproduced by a substructure, \emph{e.g.} the downturn at
$\lambda>12\,\mu\mathrm{m}$ for $\mathrm{BL\sim36\,\mathrm{m}}$,
$\mathrm{PA\sim67^\circ}$ (see Fig.~\ref{fig:circ_phys-model-compcor}).

An inhomogeneous dust distribution complicates the problem of determining
the morphology of the source significantly, when interferometric methods
are used. With the current $uv$ coverage, we are not able to say
anything about the details of the clump distribution of the source.
More data are needed to disentangle these and also to enable us to
trace how the morphology changes with the wavelength.

\subsection{Orientation and geometry of the dust distribution\label{sub:circ_orientation}}

Both direct inspection of the data (see Fig.~\ref{fig:circ_corvis})
and our modelling reveal that the source is elongated in direction
of $\mathit{PA}\sim60^\circ$. This effect is dominated by our observations
with longer baselines (UT3~--~UT4); the shorter baseline observations
(UT2~--~UT3) do not show a clear orientational preference. In our
interpretation, this means that the inner component is more disk-like,
while the outer component traces an almost spherical or very thick
toroidal component.

The projected orientation of our disk component is in very good agreement
with the orientation of several components in this AGN which are directly
related to the nuclear dust distribution: its axis coincides with
the direction of the ionisation cone, which points towards $\mathit{PA}=-44^\circ$
\citep{2000Maiolino,2000Wilson} and the bidirectional outflow observed
in CO pointing towards $\mathit{PA}=124^\circ$ and $\mathit{PA}=-56^\circ$
\citep{1998Curran}. Our measurements are hence the first direct confirmation
for the presence of a dust structure extended perpendicularly to the
outflow and the ionisation cone in an AGN without taking any previous
knowledge on the source into account.

The nucleus of Circinus is special in another feature: it displays
a very strong emission of water vapour masers. This emission was first
detected by \citet{1982Gardner}. A detailed study was performed by
\citet{2003Greenhill}: The VLBI observations of the $6_{16}-5_{23}$
transition of $\mathrm{H}_{2}\mathrm{O}$ show an S shaped locus and
a wider distribution of maser sources at the nucleus of the Circinus
galaxy. The authors attribute the sources detected to two distinct
populations: a warped, edge-on {}``accretion disk'' and a wide angle
outflow. The warped disk has an inner radius of $(0.11\pm0.02)\,\mathrm{pc}$
and extends out to $\sim0.4\,\mathrm{pc}$. These are the only observations
at the same or even better angular resolution as ours. An almost perfect
alignment is found between the outer radius of the maser disk with
a position angle of $\mathit{PA}=56\pm6^\circ$ \citep{2003Greenhill}
and our disk component (\emph{c.f.} Figs.~\ref{fig:circ_phys-model-sketch}
and \ref{fig:circ_discussion-torus-sketch}).

Because of the high axis ratio of $r_{1}=0.21$, our observations
suggest that the disk component is nearly edge-on: $i_{\mathrm{disk}}>60^\circ$.
If it were a truly edge-on disk ($i=90^\circ$), it would then have to
be warped or to have a thickness of $h/r=r_{1}$ to appear as we observe
it. When observed edge-on, the maser disk with a warp of $27^\circ$ can
easily produce the same extent as observed for our smaller component.
Considering this and the good agreement of the position angles, we
believe the maser emission originates in the dense dusty and molecular
medium associated with our disk component%
\footnote{Note that we are not able to determine an accurate position of the
MIR source, because high resolution astrometry is not possible with
MIDI. We hence \emph{assume} that the torus emission is centered on
the maser disk, although it might well be possible that there is an
offset for a slightly inclined torus.%
}.

From the present data, we cannot make any detailed statements concerning
the shape of the larger torus component, as its proportions are not
very well constrained. It must be significantly extended also in the
direction of the ionisation cone and the measurements are consistent
with an almost circular dust distribution when seen in projection.
When seen almost edge-on, this implies $h/r\sim1$. The component
can, however, not be entirely spherical but it must have an opening
angle of 90° for the ionisation cones (see Fig.~\ref{fig:circ_discussion-torus-sketch}).
From our side-on viewing position, we see a projection of this hollowed
spherical dust structure, which leads to its almost circular appearance.
More measurements at shorter baselines are necessary to provide better
constraints on the distribution of the more extended dust emission.

\subsection{Collimation of the ionisation cone and outflow\label{sub:circ_collimation}}

In the context of the two components, namely the disk and the torus,
it is interesting to address the question which component causes the
collimation of the outflow and the ionisation cone. The latter has
an opening angle of $\theta_{\mathrm{i}}=90^\circ$ and a sharp boundary
seen clearest in {[}\ion{O}{iii}{]} and H$\alpha$ \citep{2000Wilson}.
The same opening angle is found for the extended X-ray emission, overlapping
with the ionisation cone \citep{2001Smith} as well as for the CO
outflow \citep{1999Curran}. Finally, the high ionisation lines in
direction of the ionisation cone reported by \citet{2000Maiolino}
({[}\ion{Si}{vi}{]}) and \citet{2004Prieto} ({[}\ion{Si}{vii}{]})
show an even smaller opening angle. Summarised, this means that at
scales $>2\,\mathrm{pc}$ only $30\,\%$ of the solid angle are exposed
directly to radiation from the nuclear engine. This cannot be explained
by the $\cos\theta$ radiation characteristic of the accretion disk,
especially considering the sharp boundaries of the cone. On scales
less than $1\,\mathrm{pc}$, \citet{2003Greenhill} found a wide angle
outflow of maser emitters. It is supposed to cover $80$ to $90\,\%$
of the solid angle, corresponding to an opening angle of $\theta_{\mathrm{m}}=120^\circ$.
This is consistent with a collimation of the outflow by a disk with
a warp of $27^\circ$. These observations lead to the conclusion that the
large torus component extending from $\sim0.2$ to $1.0\,\mathrm{pc}$
must be responsible for the further collimation of the wide angle
outflow to the narrow one observed at larger scales (see Fig.~\ref{fig:circ_discussion-torus-sketch}).

Further support for this picture comes from the observation of molecular
hydrogen gas just below the ionisation cone on scales from $6\,\mathrm{pc}$
\citep{2006MuellerSanchez} up to $20\,\mathrm{pc}$ \citep{1998Maiolino,2000Maiolino}.
This gas most likely is part of the cooler regions of the dusty torus.
Once clouds enter the ionisation cone and are directly irradiated
by the central source, they emit H$\alpha$ and {[}\ion{Si}{vi}{]}
emission, which is observed in the ionisation cone and which is especially
strong on its southern edge. The clouds are then accelerated and entrained
by the outflow (for a more detailed discussion, see \citealt{2005Packham}).

As an alternative to the thick toroidal dust distribution in unified
models, warped disks are often put forward. Fully in line with this,
\citet{2003Greenhill} have claimed that the warped disk alone may
be responsible for the obscuration of the nuclear source and for the
collimation of the outflow in Circinus, obviating the need for a geometrically
thick dust distribution. Our observations, however, have added the
decisive component: they have proven that a geometrically thick dust
distribution is indeed present in the nucleus of Circinus. It is obvious
that the opening angle observed at larger scales is too narrow to
be produced by merely a disk with a warp of $27^\circ$. Hence the disk
component cannot be the source of the collimation for the outflow
and ionising radiation alone.

From our observations we cannot say anything about cooler dust, with
temperatures $T<100\,\mathrm{K}$, and it is likely that the structure
we observe is embedded in a larger component of cool gas and dust
extending out to the starburst on scales of $10\,\mathrm{pc}$.

\section{Conclusions\label{sec:circ_conclusions}}

We have obtained extensive interferometric observations of the nucleus
of the Circinus galaxy covering a wavelength range from $8.0$ to
$13.0\,\mu\mathrm{m}$, using the MIDI interferometer at the VLT.
Through direct analysis of the data and several steps of modelling
with increasing complexity, we have arrived at the following conclusions
for the nuclear dust distribution in this active nucleus:

\begin{enumerate}
\item The distribution of dust cannot be described by a single, simple component.
However, it appears to be distributed in two components:

\begin{enumerate}
\item a dense and warm ($T_{1}\gtrsim330\,\mathrm{K}$) disk component with
a radius of $0.2\,\mathrm{pc}$ and
\item a less dense and slightly cooler ($T_{2}\lesssim300\,\mathrm{K}$)
geometrically thick torus-like component extending out to $1\,\mathrm{pc}$
half-light radius.
\end{enumerate}
\item This disk-torus configuration is oriented perpendicularly to the ionisation
cone and to the outflow.
\item The compact, disk-like dust component coincides in orientation and
extent with the nuclear maser disk.
\item From the total energy needed to heat the dust and from the solid angle
intercepted by the dust torus, we infer a luminosity of the accretion
disk of $L_{\mathrm{acc}}\approx1\cdot10^{10}\,\mathrm{L}_{\odot}$.
This corresponds to $\sim20\,\%$ of the Eddington luminosity of the
$M_{\mathrm{BH}}=1.7\cdot10^{6}\,\mathrm{M}_{\odot}$ nuclear black
hole.
\item The properties of the larger torus component are less constrained
by the data. With an $h/r\sim1$, it is consistent with a thick, edge-on
torus which has cavities for the ionisation cone and outflow. This
component is most likely responsible for collimating the radiation
and the wide angle outflow originating in the nucleus down to the
observed opening angle of the ionisation cone and outflow of $90^\circ$
at larger distances.
\item We find strong evidence for a clumpy or filamentary dust distribution
in the torus from three lines of reasoning:

\begin{enumerate}
\item The large component only has a low effective scaling factor of $\sim20\,\%$.
\item The visibility measurements show an irregular behaviour that can be
explained by clumpiness.
\item The radial temperature dependency of the dust, $T(r)$, has a rather
shallow decrease which requires that some of the outer dust is directly
exposed to the nuclear radiation.
\end{enumerate}
\item The silicate absorption depth is less pronounced for the disk component
than for the extended component, arguing for a $10\,\mu\mathrm{m}$
silicate feature in emission in the innermost, disk-shaped regions
of the dust distribution.
\item The silicate absorption profile is consistent with that of normal
galactic dust. We hence see no evidence for dust reprocessing. This
might be connected to the moderate maximum dust temperatures, $T_{\mathrm{max}}\lesssim400\,\mathrm{K}$.
\end{enumerate}
Taking this evidence together, we conclude, that the long postulated
dust torus exists in this Seyfert 2 nucleus. It is, however, significantly
more fine structured than the doughnut-shaped models which have been
proposed as long as no spatially resolved information was available.

\bibliographystyle{aa}
\bibliography{8369circ}

\onecolumn\begin{appendix}

\section{1D model\label{sub:circ_1d-model}}

\begin{figure*}[b!]
\centering
\includegraphics{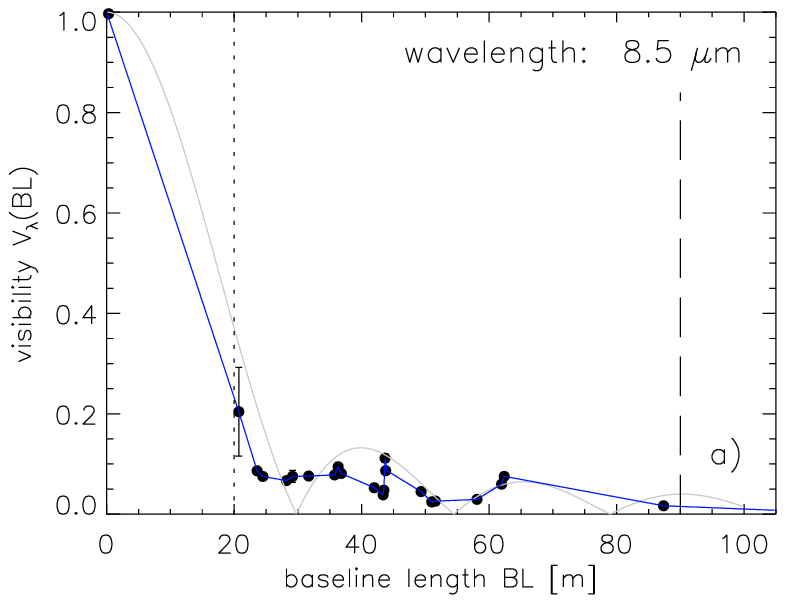}~~~\includegraphics{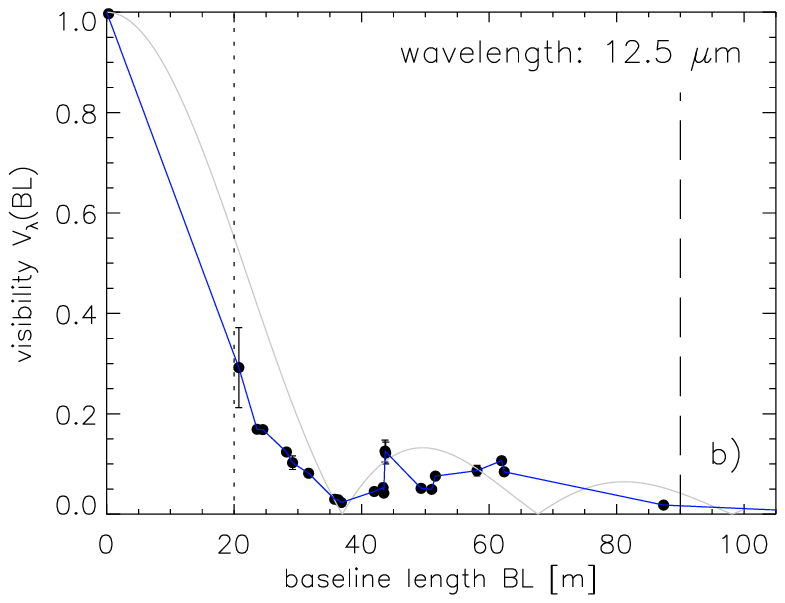}
\includegraphics{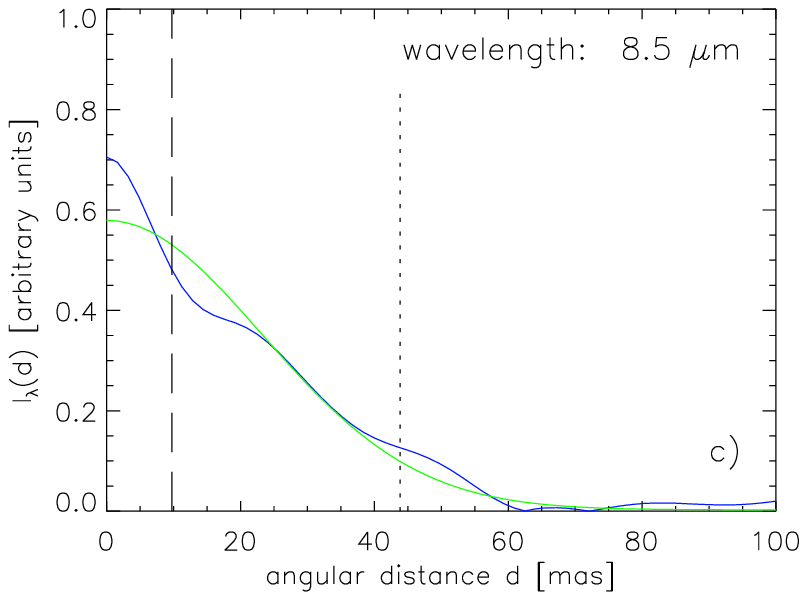}~~~\includegraphics{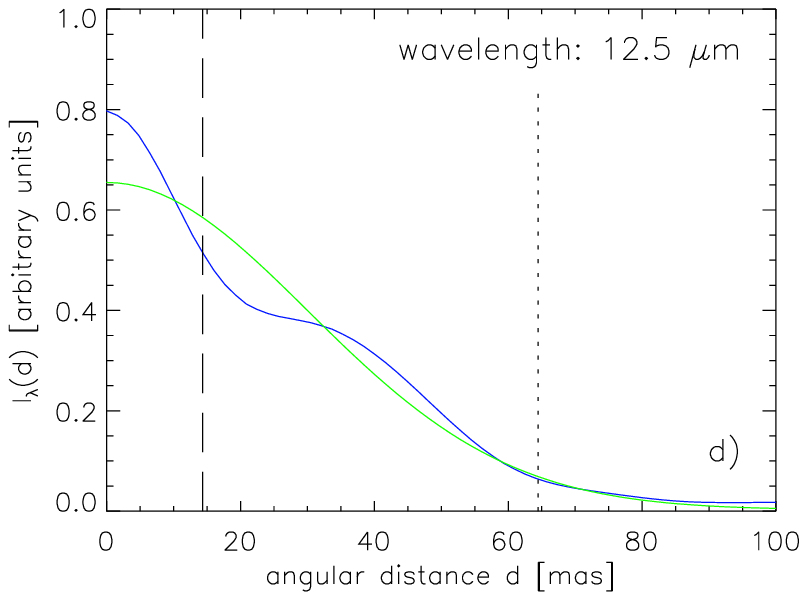}
\caption{Radial 1D analysis of the mid-infrared emission in Circinus. Top
row: Visibility as a function of baseline length for two wavelength
bins at $8.5\,\mu\mathrm{m}$ (a) and $12.5\,\mu\mathrm{m}$ (b) taking
together all position angles. The data cannot be explained by a uniform
disk model (light grey curves). The blue curve is an interpolated
visibility function derived from the data. Bottom row, (c) and (d):
For the same wavelengths, the inverse Fourier transform of the blue
curve in the top row graphs is plotted, corresponding to a brightness
distribution. Traced in green is the Gaussian distribution fitted
to the Fourier transform in order to get an estimate for the size
of the emitting source. The vertical lines indicate the angular size
range probed by the interferometric measurements.}
\label{fig:circ_1d-model-weigeltplot}
\end{figure*}

With this first step of modelling, we aim to estimate the characteristic
size of the emission region. To this end, we assumed the source to
be axisymmetric on the plane of the sky and plotted all visibility
measurements, regardless of the position angle, as a function of the
baseline length $\mathit{BL}$: $V_{\lambda}(\mathit{BL})$. This
can be done independently for all wavelength bins measured by MIDI.
Examples for such a plot are shown for two wavelength bins at $8.5\,\mu\mathrm{m}$
and $12.5\,\mu\mathrm{m}$ in the top row of Fig.~\ref{fig:circ_1d-model-weigeltplot}.%
 The visibility estimates show a decrease towards longer baselines
from $V=1$ (\textbf{$\mathit{BL}=0\,\mathrm{m}$}) and an irregularly
oscillating behaviour for \textbf{$\mathit{BL}>35\,\mathrm{m}$}.
At first glance, this visibility function is reminiscent of that for
a uniform disk, which is a Bessel function of the first kind of order
1: $\left|J_{1}(\mathit{BL})\right|$. The direct comparison in Fig.~\ref{fig:circ_1d-model-weigeltplot}
reveals, however, significant discrepancies between such a Bessel
function and the data. Additionally, our data shows no evidence for
the $2\pi$ phase jumps that occur at the zero crossings of the Bessel
function. We thus conclude that a uniform disk model cannot explain
the data.

We chose an alternative approach to analyse the data: First we interpolated
$V_{\lambda}(\mathit{BL})$ (blue curves in the top row of Fig.~\ref{fig:circ_1d-model-weigeltplot})
and then applied an inverse Fourier transform to this visibility function
for each wavelength bin: $I_{\lambda}(d)=\mathcal{F}_{\mathit{BL}}\left[V_{\lambda}(\mathit{BL})\right](d)$.
This reconstruction method gives a first order estimate of the radial
brightness distribution of the nucleus $I_{\lambda}(d)$, where $d$
is the angular distance from the centre. The shape of the visibility
function $V_{\lambda}(BL)$ is only well determined for $20\,\mathrm{m}<\mathit{BL}<90\,\mathrm{m}$.
This in turn implies that we have probed angular sizes $d$ between
$10$ and $70\,\mathrm{mas}$, considering the entire wavelength range
of the MIDI observations. These characteristic sizes in the Fourier
and the real domain are indicated by vertical dashed and dotted lines
in Fig.~\ref{fig:circ_1d-model-weigeltplot}.

\begin{figure}
\centering
\includegraphics{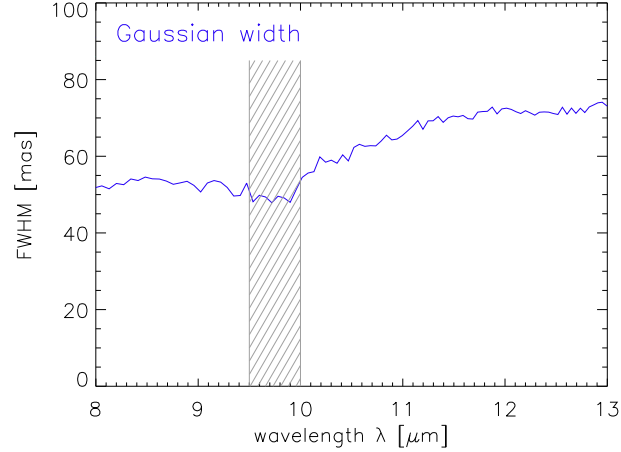}
\caption{Characteristic size of the central dust emission in Circinus as a
function of wavelength: the figure shows the $\mathit{FWHM}$ of the
Gaussian distribution fitted to the radial profile of the brightness
distribution of the Circinus nucleus (see Fig. \ref{fig:circ_1d-model-weigeltplot}).
This first estimate for the effective size of the nuclear dust emission
shows a slow increase with wavelength from $50$ to $70\,\mathrm{mas}$.
The area affected by atmospheric absorption are hatched in grey.}
\label{fig:circ_1d-model-widths}
\end{figure}

The reconstructed brightness distribution is composed of two components
plus a small fraction of extended flux at distances of more than $100\,\mathrm{mas}$
of the centre. The extended component is fully resolved by our interferometric
set-up, while our measurements probe the properties of the other two
components. One should bear in mind that we assumed axisymmetry for
this analysis and already from the qualitative discussion of the data
in Sect.~\ref{sec:circ_results} we know that this is a crude simplification.
We therefore used the reconstructed brightness distribution only to
extract the effective size of the emission region in the mid-infrared
and traced its evolution depending on the wavelength. This was achieved
by fitting a Gaussian distribution to the reconstructed radial profiles
at every wavelength bin. This Gaussian is shown in green in the bottom
row of Fig.~\ref{fig:circ_1d-model-weigeltplot}. The full width
at half maximum (FWHM) of this Gaussian distribution is shown as a
function of wavelength in Fig.~\ref{fig:circ_1d-model-widths}.%
 We attribute the small dip in the size between $9.5$ and $10.0\,\mu\mathrm{m}$
to artefacts coming from the ozone feature (hatched area in Fig.~\ref{fig:circ_1d-model-widths}).
From shorter to longer wavelengths, the size of the flux distribution
increases slightly from $50$ to $75\,\mathrm{mas}$. This indicates
that the emission at longer wavelengths is more extended than at short
wavelengths: the temperature of the dust decreases with distance to
the nucleus, but apparently only slowly.

These effective size estimates again demonstrate the need for interferometric
observations, as they are the only way to reach these resolutions
in the infrared.

\section{2D geometrical model\label{sub:circ_geo-model}}

This second model is a two-dimensional, purely geometrical model,
aimed to reconstruct the brightness distribution of the source from
the full two-dimensional visibility data set. The model consists of
two concentric, elliptical Gaussian distributions. It was motivated
by evidence for two components in the one-dimensional analysis and
by the dependence of the visibilities on the position angle which
imply an elongated source (Sect.~\ref{sub:circ_interferometric-data}
and there especially Fig.~\ref{fig:circ_corvis}). The brightness
distribution has the following functional form:%
\footnote{To be more descriptive, we use the FWHM, $\Delta_{n}=2\sqrt{2\ln2}\,\sigma_{n}$,
in our definition of the Gaussian distribution instead of the sigma,
$\sigma_{n}$, in the standard definition.%
}

\begin{eqnarray}
I(\lambda,\alpha,\delta)=f(\lambda) & \cdot & \exp\left(-4\ln2\cdot\left[\left({\textstyle \frac{\alpha\cos\phi(\lambda)+\delta\sin\phi(\lambda)}{r_{1}(\lambda)\cdot\Delta_{1}(\lambda)}}\right)^{2}+\left({\textstyle \frac{\alpha\sin\phi(\lambda)-\delta\cos\phi(\lambda)}{\Delta_{1}(\lambda)}}\right)^{2}\right]\right)\nonumber \\
 & + & \exp\left(-4\ln2\cdot\left[\left({\textstyle \frac{\alpha\cos\phi(\lambda)+\delta\sin\phi(\lambda)}{r_{2}(\lambda)\cdot\Delta_{2}(\lambda)}}\right)^{2}+\left({\textstyle \frac{\alpha\sin\phi(\lambda)-\delta\cos\phi(\lambda)}{\Delta_{2}(\lambda)}}\right)^{2}\right]\right)\label{eq:circ_geo-model-intens}\end{eqnarray}

Here, $\alpha$ and $\delta$ are the position on the sky with respect
to the centre of the distribution. This two component model has a
total of six free parameters: the sizes of the two Gaussians, given
by their full width at half maximum $\Delta_{1}(\lambda)$ and $\Delta_{2}(\lambda)$;
their oblateness, represented by the axis ratio $r_{1}(\lambda)$
and $r_{2}(\lambda)$; the position angle $\phi(\lambda)$ of the
major axis with respect to north assumed to be the same for both components;
and the relative intensity of the two components quantified $f(\lambda)$.

\begin{figure*}
\centering
\includegraphics{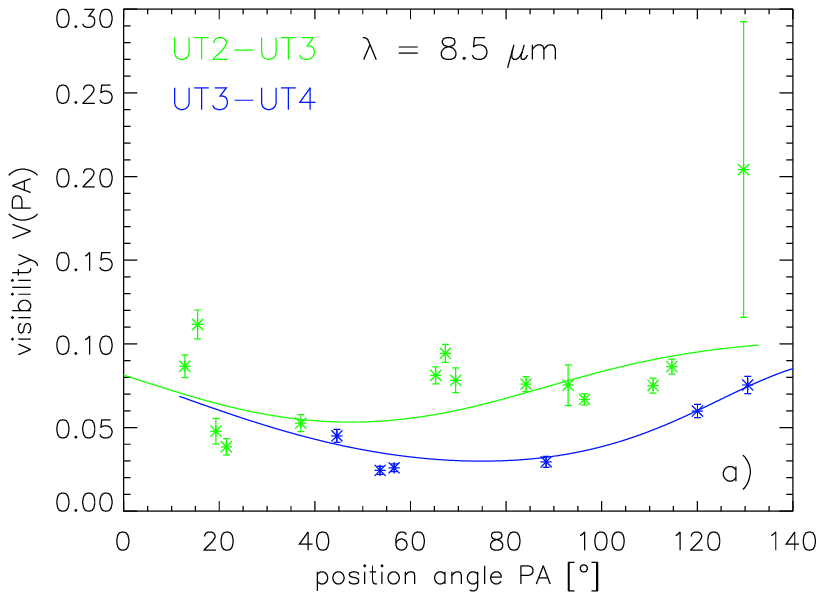}~~~\includegraphics{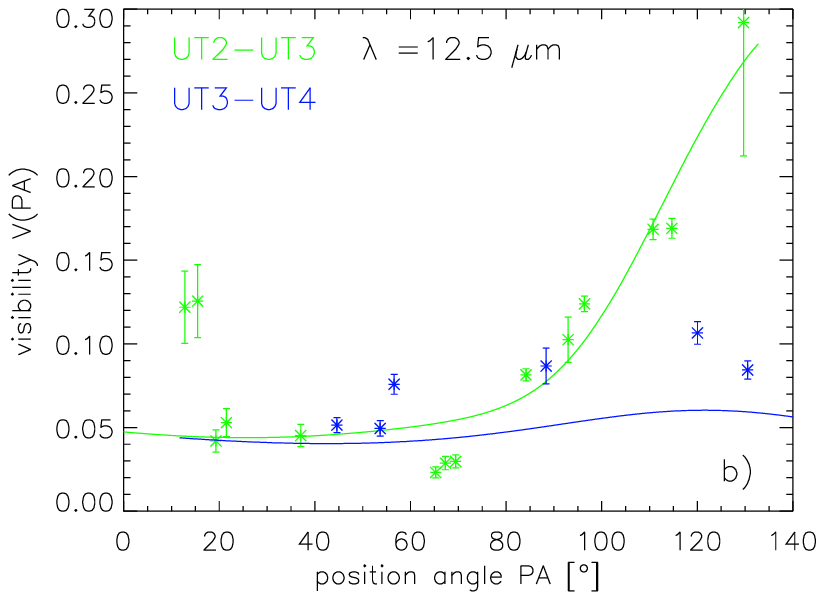}
\caption{Comparison of the visibilities between the 2D geometrical model (two
elliptical Gaussian distributions) and the measured data for two wavelengths:
(a) $8.5$ and (b) $12.5\,\mu\mathrm{m}$. The data are plotted with
asterisks; the modelled visibilities are the continuous lines. The
two baselines shown here are plotted in two different colours, UT2~--~UT3
in green (grey) and UT3~--~UT4 in blue (black).}
\label{fig:circ_geo-model-compare}
\end{figure*}

\begin{figure*}
\centering
\includegraphics{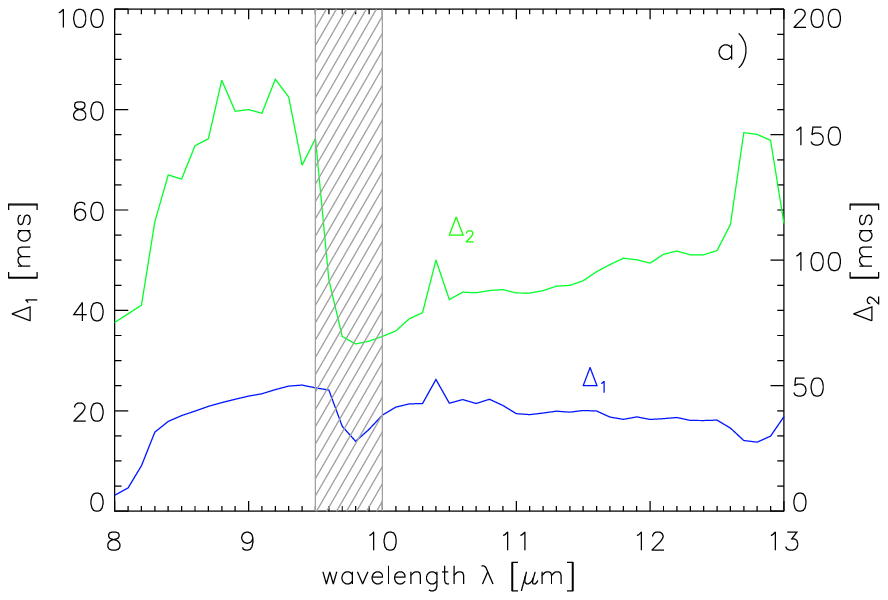}~~~\includegraphics{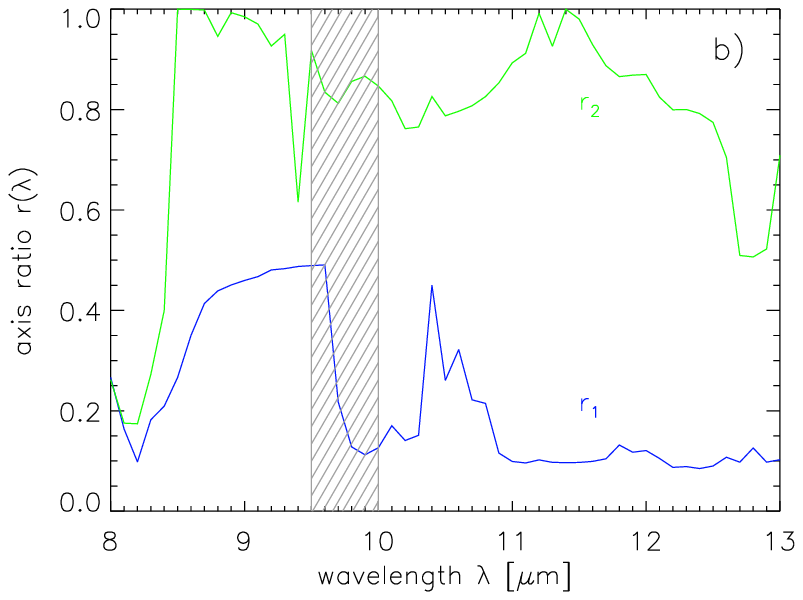}
\includegraphics{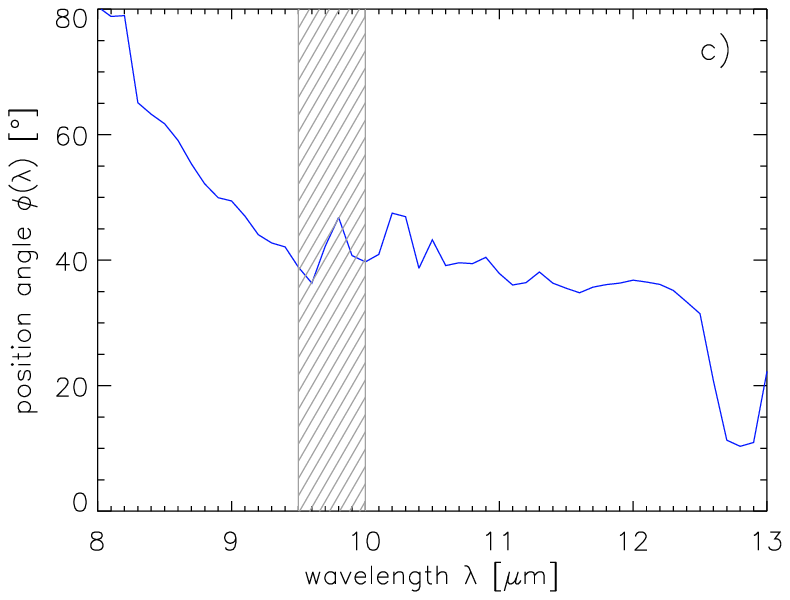}~~~\includegraphics{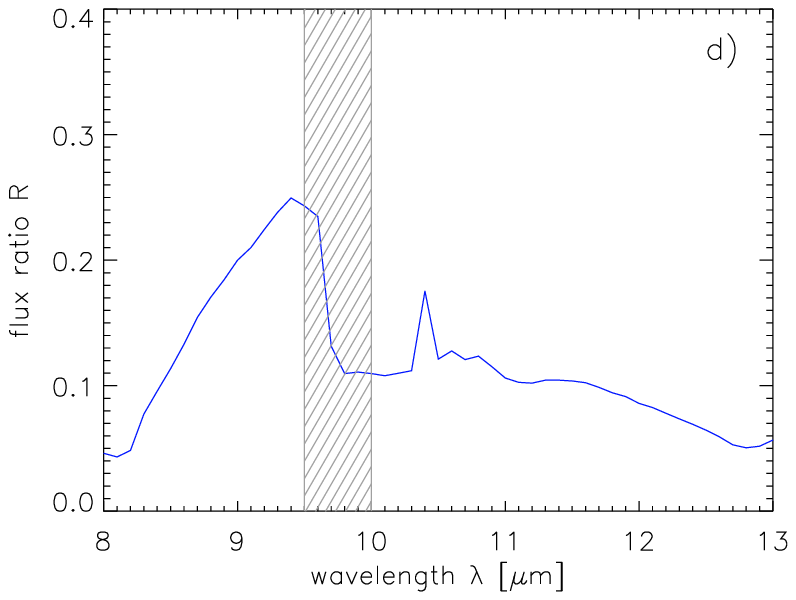}
\caption{Wavelength dependence of the parameter estimates for the two-dimensional
geometric model. The six parameters are the two $\mathit{FWHM}$ of
the Gaussian components (a), the respective axis ratios $r$ (b),
the position angle $\phi$ of the elliptical distribution (c) and
the ratio $R$ of the flux in the compact over the extended component
(d). The areas affected by atmospheric absorption are hatched in grey.}
\label{fig:circ_geo-model-params}
\end{figure*}

\begin{figure*}
\centering
\includegraphics{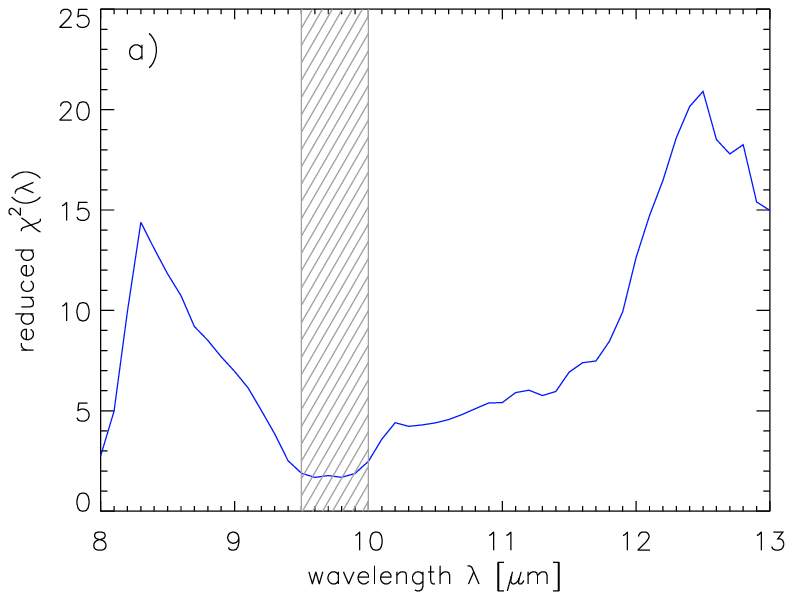}~~~\includegraphics{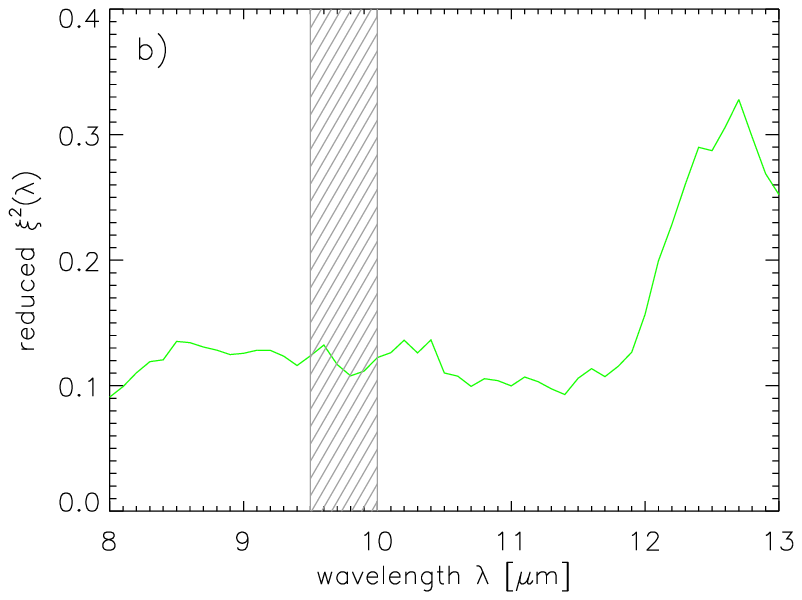}
\caption{$\chi^{2}$ (a) and $\xi^{2}$ (b) for the fit of the two component
Gaussian model to the visibilities. The areas hatched in grey are
affected by atmospheric absorption.}
\label{fig:circ_geo-model-chisq}
\end{figure*}

These parameters were then optimised using the following prescription:
for a given wavelength $\lambda$ the flux distribution $I_{\lambda}(\alpha,\delta)=I(\lambda,\alpha,\delta)$
was generated according to the set of parameters. Then the Fourier
transform $\mathcal{V_{\lambda}}(u,v)=\mathcal{F}_{\alpha,\delta}\left[I_{\lambda}(\alpha,\delta)\right](u,v)$
was calculated using a fast Fourier transform (FFT). The result corresponds
to the visibility distribution in the $uv$ plane. The visibility
values for the model were extracted from this plane at the positions
$u_{i}=\mathit{BL}_{i}\cdot\sin\mathit{PA}_{i}$, $v_{i}=\mathit{BL}_{i}\cdot\cos\mathit{PA}_{i}$,
using the baseline information from our observations $i$ consisting
of the baseline length $\mathit{BL}_{i}$ and the position angle $\mathit{PA}_{i}$.
Finally, the modelled visibilities $V_{i}^{\mathrm{mod}}(\lambda)=\left|\mathcal{V_{\lambda}}(u_{i},v_{i})\right|$
were compared to the observed ones $V_{i}^{\mathrm{obs}}(\lambda)$.
The discrepancy was weighted with the measurement errors $\sigma_{V_{i}^{\mathrm{obs}}}(\lambda)$
to obtain the $\chi^{2}$: \begin{equation}
\chi^{2}(\lambda)={\displaystyle \sum_{i=1}^{21}\left(\frac{V_{i}^{\mathrm{obs}}(\lambda)-V_{i}^{\mathrm{mod}}(\lambda)}{\sigma_{V_{i}^{\mathrm{obs}}}(\lambda)}\right)^{2}}.\end{equation}
 This $\chi^{2}$ was minimised using the Levenberg-Marquardt least-squares
minimisation algorithm. This process was performed independently for
the 51 wavelength bins measured by MIDI between $8$ and $13\,\mu\mathrm{m}$.
A comparison of the model to the data is shown for $8.5$ and $12.5\,\mu\mathrm{m}$
in Fig.~\ref{fig:circ_geo-model-compare}.%
 This plot is the same as Fig.~\ref{fig:circ_phys-model-compare},
except that here the visibilities are compared instead of the correlated
fluxes. The resulting parameter dependencies are shown in Fig.~\ref{fig:circ_geo-model-params}.%
 To have more practical parameters, the factor $f$ was replaced by
the ratio $R$ of the flux in the compact component to the flux in
the extended component. The reduced $\chi^{2}$ (that is $\chi^{2}(\lambda)/N_{\mathrm{free}}$
with $N_{\mathrm{free}}=21-6=15$ degrees of freedom), shown in the
left panel of Fig.~\ref{fig:circ_geo-model-chisq}, varies strongly
for different wavelengths.%
 The fits are best in the middle of the wavelength range. Towards
the edges of the N band, the fits get significantly worse. This is
mainly due to the weighting of the discrepancy between model and data
using the errors of the data $\sigma_{i}^{\mathrm{obs}}(\lambda)$.
Due to the bad signal-to-noise ratio for low fluxes, the errors of
the visibility are of the same order, or even larger within the silicate
absorption feature, the atmospheric ozone feature, as well as at the
edges of the N band. The larger errors lead to a lower $\chi^{2}$
value. When considering a modified estimate for the quality of the
fit, \begin{equation}
\xi^{2}(\lambda)={\displaystyle \sum_{i=1}^{21}\left(\frac{V_{i}^{\mathrm{obs}}(\lambda)-V_{i}^{\mathrm{mod}}(\lambda)}{V_{i}^{\mathrm{mod}}(\lambda)}\right)^{2}},\end{equation}
the discrepancy is of the same order, except at $\lambda>12\,\mu\mathrm{m}$.
The reduced $\xi^{2}$, $\xi^{2}(\lambda)/N_{\mathrm{free}}$, is
shown in the right panel of Fig.~\ref{fig:circ_geo-model-chisq}.

With the high values of the reduced $\chi^{2}$, the fit is comparatively
bad. This is partly due to an overly optimistic estimate of the errors
of the visibilities. The errors only represent the relative errors
within one measurement and do not reflect the systematic errors between
two measurements at different epochs. More importantly, we see in
the bad fit quality evidence for a more complex structure of the source.
This will be discussed in Sect.~ \ref{sub:circ_physical-model}.

Although no clear dependency of the parameters with wavelength is
identifiable, a few general trends are present. The smaller component
has a rather constant size of approximately $\Delta_{1}\sim20\,\mathrm{mas}$
and a rather high axis ratio of $r_{1}=0.2$. The position angle has
a preferred orientation of $\mathit{PA}\sim40^\circ$, which is more or
less perpendicular to the ionisation cone. Towards both ends of the
spectral range sampled ($8.0\,\mu\mathrm{m}<\lambda<9.0\,\mu\mathrm{m}$
and $12.5\,\mu\mathrm{m}<\lambda<13.0\,\mu\mathrm{m}$), the position
angle deviates significantly from this value. A more regular behaviour
of the parameters is found in the range from $11.0$ to $12.5\,\mu\mathrm{m}$.
Here also the intensity ratio stabilises at $R\sim0.1$ and the size
of the second component grows linearly with increasing wavelength.
The large component hence comprises the bulk of the emission. The
ellipticity of the larger component is relatively small, with $r_{2}\sim0.75$
for large parts of the spectral range, again exceeding these margins
at the edges of the spectrum sampled. Some of the parameters show
a certain reciprocal influence: decreasing the inclination of the
smaller component between $8.2$ and $9.5\,\mu\mathrm{m}$ leads to
a larger emitting region and hence to a larger flux, which has an
influence on the flux ratio between the two components and the size
of the larger component. We hence do not see the decrease of ellipticity
of the small component to be significant but rather as an artifact
of the ambiguity of the data. Especially the ellipticity, the size
and the flux ratio of the two components show a certain degree of
degeneracy. We found that, by fixing some of the parameters to average
values (\emph{e.g.} $\mathit{PA}$ to $40^\circ$ or $r_{1}$ to $0.2$),
the quality of the fit for the individual wavelengths does not decrease
significantly. At the same time, however, the behaviour of the other
parameters, which are then adjusted, looks steadier. We did not follow
this path of analysis any further but decided to apply a physical
model instead (see Sect.~\ref{sub:circ_physical-model}). The overall
erratic behaviour of the parameters and the quality of the fit, however,
show that two Gaussians are not a good fit and that there are substantial
differences at different wavelength bins. Nonetheless, we find that
other models of the same simplicity produce no better fits. A more
detailed discussion of alternative models is be given in Sect.~\ref{sub:circ_model-selection}.

\section{Comparison of the full data set to the physical model\label{sec:appendix}}

\begin{figure}[H]
\begin{center}\includegraphics[bb=22bp 31bp 555bp 788bp,
  scale=0.81]{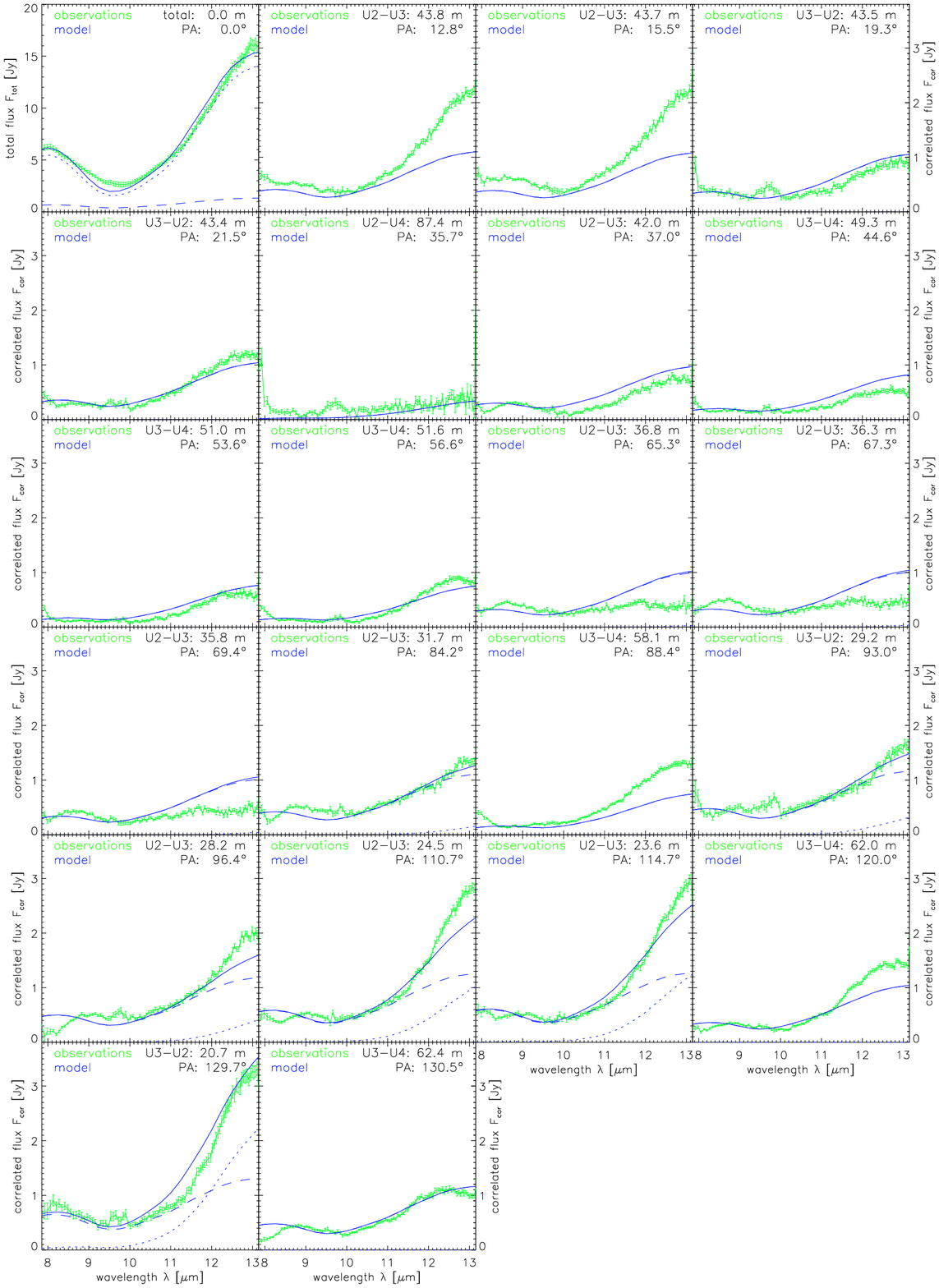}\end{center}
\caption{Comparison of the dispersed correlated fluxes (green/gray) to the
model fluxes for the physical model fit (fit2, blue/black) in the
entire wavelength range from $8$ to $13\,\mu\mathrm{m}$. The dashed
line represents the flux contribution of the small disk component,
while the dotted line is the contribution of the large torus component.\label{fig:circ_phys-model-compcor}}
\end{figure}

\end{appendix}
\end{document}